\documentclass[letterpaper,11pt,onecolumn]{article}
\usepackage[utf8]{inputenc}
\usepackage[T1]{fontenc}
\usepackage[letterpaper,margin=1in]{geometry}
\usepackage{setspace}
\usepackage{amsmath,amsfonts,amssymb}
\usepackage{physics,mathrsfs,bbold,bm}
\usepackage{graphicx}

\usepackage{xcolor}
\usepackage{array,booktabs,longtable,multirow,makecell,threeparttable}
\usepackage{floatrow}
\usepackage[font=small,labelfont=bf]{caption}
\usepackage{soul}
\usepackage[version=4]{mhchem}
\usepackage[journal=jctcce,articletitle=true,doi=true,maxauthors=0]{achemso}
\usepackage[hidelinks]{hyperref}
\usepackage{xurl}
\usepackage{authblk}

\usepackage{lineno}
\let\oldalign\align
\let\oldendalign\endalign
\renewenvironment{align}
{\linenomathNonumbers\oldalign}
{\oldendalign\endlinenomath}
\AtBeginEnvironment{table}{\small\singlespacing}
\AtBeginEnvironment{longtable}{\small\singlespacing}
 
\newcommand{\articletitle}{Toward Long-Range Correlation Energies from Charge Fluctuations: Incorporating Exact Exchange into ACKS2\texorpdfstring{$\omega$}{omega}}
 \title{\articletitle}
\hypersetup{pdftitle={\articletitle}}
\author[1,2]{YingXing Cheng\thanks{Corresponding author: \href{mailto:chengyx@ms.xjb.ac.cn}{\nolinkurl{chengyx@ms.xjb.ac.cn}}}}
\hypersetup{pdfauthor={YingXing Cheng}}
\affil[1]{Research Center for Crystal Materials; CAS Key Laboratory of Functional Materials and Devices for Special Environmental Conditions; Xinjiang Key Laboratory of Functional Crystal Materials; Xinjiang Technical Institute of Physics and Chemistry, Chinese Academy of Sciences, 40-1 South Beijing Road, Urumqi 830011, China.}
\affil[2]{Center of Materials Science and Optoelectronics Engineering, University of Chinese Academy of Sciences, Beijing 100049, China.}
 \date{}

\makeatletter
\begin{document}
\nolinenumbers
\maketitle

\begin{abstract}
Long-range electronic correlation can be represented in terms of coupled fluctuations of atomic charges and higher multipoles, motivating a density-response description beyond local dipolar polarization, particularly in low-dimensional, conducting, and small-gap systems where collective electronic fluctuations become important.
We extend the frequency-dependent polarizable force field ACKS2$\omega$ to electronic-structure references containing exact exchange by deriving both the static and additional frequency-dependent exchange contributions, with particular attention to HFsrPBE, which combines long-range Hartree--Fock exchange with short-range PBE exchange and correlation.
Numerical validation of ACKS2$\omega$ using dipole polarizabilities and $C_6$ dispersion coefficients for the 27 molecules of the TS27 set, with HF, PBE, B3LYP, and HFsrPBE as the underlying electronic-structure references, shows that including the exchange contributions and enriching the response basis improve agreement with the corresponding orbital-space linear-response calculations, although finite-response-basis errors remain molecule dependent.
Long-range correlation energies evaluated through the adiabatic-connection fluctuation-dissipation theorem using the ACKS2$\omega$ parameters reproduce the corresponding HFsrPBE orbital-space results for small test molecules when the response basis is made complete within the chosen orbital basis, while a local multipole decomposition resolves the finite-basis energies into monopole--monopole, mixed, dipole--dipole, and remaining higher-multipole contributions, providing a route to evaluating long-range correlation energies in terms of charge fluctuations within atom-condensed models.
 \end{abstract}

\section{Introduction}
\label{sec:introduction}

London dispersion is a manifestation of long-range dynamical electron correlation and contributes to noncovalent binding in molecules and materials.
It arises from correlated fluctuations of the electronic density and persists even between neutral fragments without permanent electric multipole moments.\cite{London1930,Casimir1948}
Several electronic-structure approaches can be used to describe such interactions.
Post-Hartree--Fock wave-function methods provide one route by treating electron correlation explicitly.\cite{Moller1934,Raghavachari1989,Jeziorski1994,Jurecka2006,Takatani2010,Rezac2011}
M{\o}ller--Plesset perturbation theory includes dispersion already at second order.\cite{Moller1934,Jurecka2006}
Higher-level coupled-cluster methods, particularly CCSD(T), are widely used to obtain benchmark intermolecular interaction energies.\cite{Raghavachari1989,Takatani2010,Rezac2011}
Symmetry-adapted perturbation theory (SAPT) provides a complementary description in which dispersion appears explicitly as one component of the intermolecular interaction alongside electrostatics, exchange, and induction.\cite{Jeziorski1994}
High-level wave-function calculations have consequently been used to construct benchmark data sets for noncovalent interaction energies and to assess more approximate electronic-structure methods.\cite{Jurecka2006,Takatani2010,Rezac2011}
These approaches can provide detailed and highly accurate descriptions, although their computational cost, particularly for high-level coupled-cluster and SAPT calculations, limits their practical application as molecular size increases.

Density-functional theory (DFT) provides a computationally attractive alternative for larger systems, but local and semilocal exchange-correlation (XC) approximations do not recover the correct asymptotic dispersion interaction between well-separated fragments.
Several strategies have therefore been developed to account for long-range dispersion within DFT and can be broadly classified into three categories.
The first supplements a conventional XC functional with an explicit dispersion correction, as in atom-pair DFT-D and density-dependent schemes.\cite{Grimme2010,Tkatchenko2009}
The second incorporates nonlocal long-range correlation directly into the XC functional, as in the vdW-DF and VV10 families of nonlocal density functionals.\cite{Dion2004,Lee2010,Vydrov2010}
The third obtains the long-range correlation energy from the electronic response through the adiabatic-connection fluctuation-dissipation theorem (ACFDT) and can be combined with range separation.\cite{Toulouse2009,Toulouse2010}
ACFDT relates the correlation energy to density fluctuations through the frequency-dependent density response integrated over interaction strength.\cite{Langreth1975,Gunnarsson1976}
In this work, we focus on the third category and, in particular, on representing the long-range electronic response and correlation energy in an atom-condensed form.

Among these DFT strategies, atom-pair dispersion corrections provide a particularly simple and widely used description.
They represent dispersion through damped atom-pair terms with coefficients that depend on the local chemical environment.\cite{Grimme2010,Tkatchenko2009}
The connection between these coefficients and electronic response follows from the Casimir--Polder relation, according to which the leading isotropic $C_6$ coefficient is determined by the dipole polarizabilities of the two fragments evaluated at imaginary frequencies.\cite{Casimir1948}
In a distributed multipole description, atoms or groups of atoms are treated as polarizable sites, and higher-multipole responses contribute to higher-order terms in the intermolecular interaction.\cite{Misquitta2006}
Such site-based representations are convenient for molecular force fields and atom-based dispersion corrections.\cite{Misquitta2006,Grimme2010,Tkatchenko2009}
Their simplest pairwise forms, however, do not include the full coupling among electronic fluctuations on different sites.\cite{Tkatchenko2012,Ambrosetti2014,Dobson2014}

Departures from simple atom-pair additivity can be organized using Dobson's classification of dispersion nonadditivity.\cite{Dobson2014}
Type-A effects describe changes in atomic response caused by the chemical environment, type-B effects arise from electrodynamic coupling among polarizable sites, and type-C effects involve spatially extended charge fluctuations associated with strong electronic delocalization or small electronic gaps.\cite{Dobson2006,Dobson2014}
The latter can qualitatively alter the asymptotic distance dependence of dispersion in low-dimensional conducting systems.\cite{Dobson2006,Dobson2014}

Several widely used dispersion models incorporate type-A effects through atomic dispersion coefficients that depend on the chemical environment.
Examples include the Tkatchenko--Scheffler (TS) method,\cite{Tkatchenko2009} DFT-D3 and DFT-D4,\cite{Grimme2010,Caldeweyher2019} and the exchange-hole dipole moment (XDM) model.\cite{BeckeJohnson2005DFT}
DFT-D3 and DFT-D4 can additionally include an Axilrod--Teller--Muto (ATM) three-body contribution.\cite{Grimme2010,Caldeweyher2019}
Analogous three-body terms have also been derived for XDM.\cite{OteroXDM2013}
The three-body ATM term represents the lowest-order type-B contribution, but higher-order many-center terms are required to describe the full electrodynamic screening associated with type-B behavior.\cite{Dobson2014}
The many-body dispersion (MBD) model treats these effects more systematically by coupling atomic dipole oscillators and thereby including collective polarization and screening to all orders.\cite{Tkatchenko2012,Ambrosetti2014,Dobson2014}
Because standard MBD remains based on local dipolar fluctuations, however, it does not explicitly include the long-range intersite charge fluctuations required for type-C behavior.\cite{DobsonAmbroselli2023}

More generally, charge flow describes redistribution of electron population between atomic regions and is distinct from local dipole polarization, which describes deformation of the density within a region.\cite{Misquitta2006,Cheng2023}
Charge flow can also couple to dipole and higher-multipole polarization, so its contribution is not restricted to charge--charge response.
Fluctuating-charge extensions can incorporate these effects, as illustrated by MBD+C, which couples fluctuating charges and dipoles and recovers type-C asymptotic behavior in low-dimensional metallic systems.\cite{DobsonAmbroselli2023}
In finite molecular systems, the atom-condensed response has also been used to analyze electron delocalization and resonance.\cite{SablonDelocalization2010}
Charge flow can make substantial contributions to molecular and cluster polarizabilities\cite{Mei2015,Jackson2016} and to molecular dispersion coefficients.\cite{Cheng2023}
These observations motivate a response description that includes both charge flow and local polarization.

Nonlocal distributed polarizabilities provide a direct way to represent both local polarization and intersite charge flow.
In the hydrogen-chain models studied by Misquitta and co-workers, the nonlocal charge-flow polarizabilities become increasingly long-ranged as the electronic gap decreases and account for both the enhancement and the unusual distance dependence of the dispersion interaction.\cite{Misquitta2010,Misquitta2014}
Distributed polarizabilities can therefore describe the collective charge fluctuations associated with type-C behavior when the required nonlocal response is included.
They can be constructed from the interacting molecular response using constrained density fitting or atoms-in-molecules (AIM) partitioning.\cite{Misquitta2006,Cheng2023}
The main practical difficulty is that the interacting response must first be calculated at the electronic-structure level, which can become computationally demanding for large systems and for frequency-dependent response.

ACKS2 was originally formulated within Kohn--Sham DFT and provides a compact description of the interacting electronic response through coupled density and auxiliary-potential variations.\cite{Verstraelen2013,Verstraelen2014,Gutlein2019,Gutlein2020}
ACKS2$\omega$ extends this framework to frequency-dependent response and allows both charge flow and local multipole polarization to be described.\cite{Cheng2022,Cheng2023}
These features make ACKS2$\omega$ a natural starting point for constructing an atom-condensed treatment of long-range correlation within ACFDT.

To combine such an explicit long-range correlation treatment with DFT, the correlation evaluated explicitly must be separated from that already included in the density functional.
Range-separated DFT provides a natural framework for this purpose by dividing the electron--electron interaction into complementary short- and long-range parts.\cite{Toulouse2009,Toulouse2010}
In this work, we employ an HFsrPBE reference, which combines long-range Hartree--Fock (HF) exchange with short-range PBE exchange and correlation, and treat the remaining long-range correlation within an ACFDT framework.\cite{Goll2005,Toulouse2010,Rebolini2013b}

Applying ACKS2$\omega$ to such a reference introduces an additional challenge because exact exchange is nonlocal and depends on the one-particle density matrix.
Previous applications of ACKS2$\omega$ focused mainly on semilocal XC functionals, for which the response could be described without explicitly accounting for the variation of a nonlocal exchange operator.\cite{Cheng2022,Cheng2023}
For HF and hybrid references, however, the exchange contribution changes as the electronic state responds to an external perturbation and must therefore be incorporated consistently into the response model.\cite{Rebolini2013b}
This issue is also important for long-range correlation methods based on the random-phase approximation (RPA).
In RPA with exchange (RPAx), the exchange interaction additionally enters the coupled electronic response.\cite{Toulouse2010,Angyan2011}

In this work, we extend ACKS2$\omega$ to exchange-containing references and develop an atom-condensed framework for evaluating long-range correlation energies within ACFDT.
We apply the formulation to an HFsrPBE range-separated reference and evaluate the long-range correlation using the direct random-phase approximation, variant I (dRPA-I), and the random-phase approximation with exchange, variant I (RPAx-I).\cite{Toulouse2010,Angyan2011}
Frequency-dependent polarizabilities and $C_6$ coefficients are also examined for semilocal, hybrid, range-separated, and HF references to assess the resulting ACKS2$\omega$ description.
The atom-condensed representation further allows the long-range correlation energy to be analyzed in terms of charge redistribution and local polarization.

The remainder of this paper is structured as follows.
The theoretical formulation is presented in Sec.~\ref{sec:theory}, followed by the computational details in Sec.~\ref{sec:computational-details}, the numerical results in Sec.~\ref{sec:numerical-validation}, the discussion in Sec.~\ref{sec:discussion}, and the conclusions in Sec.~\ref{sec:summary}.
Atomic units are used throughout unless noted otherwise.
 \section{Theory}
\label{sec:theory}
In this work, we follow the notation of Refs.~\citenum{Cheng2022,Cheng2023} throughout the derivations.

\subsection{ACKS2 and ACKS2\texorpdfstring{$\omega$}{omega} working equations}
\label{subsec:acks2-review}

ACKS2 is a first-principles-based model of electronic polarization that extends charge-equilibration descriptions through a second-order expansion of the electronic energy about a reference state.\cite{Verstraelen2013,Verstraelen2014}
Within a chosen atomic partition, it describes molecular polarization through both charge flow, corresponding to redistribution of electron population between atomic regions, and local polarization, corresponding to deformation of the electron density within those regions.
The model parameters have defined physical roles and can be calculated from a chosen electronic-structure reference, linking the atom-condensed description to the underlying molecular response.\cite{Verstraelen2014}
Although ACKS2 was originally formulated within Kohn--Sham DFT, the model can be defined for a more general variational reference.\cite{Verstraelen2014}
Here we summarize the working equations needed for the frequency-dependent exchange extension.

In ACKS2, the induced electron-number density, $\Delta\rho$, is described together with an auxiliary potential, $\Delta u$, in response to an external perturbation $\Delta v$.
The auxiliary electronic system describes how the reference orbitals redistribute electron density in response to the perturbation, whereas the explicit interaction hardness describes how the explicitly treated electron--electron interactions couple these density fluctuations.\cite{Verstraelen2014,Cheng2022}
Their coupled response determines the molecular polarization at fixed total electron number.
The frequency-dependent extension, ACKS2$\omega$, applies a similar description to oscillating perturbations.\cite{Cheng2022}
Accordingly, we expand the induced density and auxiliary potential in $M$ density basis functions $f_m$ and $N$ potential basis functions $g_n$, respectively:
\begin{align}
    \Delta\rho(\bm r,\omega)
    &=\sum_{m=1}^{M}C_m(\omega)f_m(\bm r), \nonumber \\
    \Delta u(\bm r,\omega)
    &=\sum_{n=1}^{N}U_n(\omega)g_n(\bm r).
    \label{eq:review-acks2-expansions}
\end{align}
The corresponding expansion coefficients, $C_m$ and $U_n$, depend on frequency, whereas the basis functions remain fixed.
Setting $\omega=0$ recovers the static ACKS2 expansions.\cite{Verstraelen2014}
Choosing atomic population and multipole functions gives an atom-condensed representation of the response, although the density and potential bases may be chosen independently.
In static ACKS2, complete density and potential response bases recover the linear response of the chosen electronic-structure theory.\cite{Verstraelen2014}
For the frequency-dependent exchange extension considered here, the corresponding density-matrix response and exchange contributions must also be specified, as described in Sec.~\ref{subsec:main-exchange-response}.

With these basis expansions, the ACKS2$\omega$ working equations take the following matrix form\cite{Cheng2022}
\begin{align}
 \begin{bmatrix}
 -\bm\eta^{\mathrm{exp}}(\omega) & \bm O & \bm D\\
 \bm O^{\mathsf T} & -\bm\chi^{\mathrm{imp}}(\omega) & \bm0\\
 \bm D^{\mathsf T} & \bm0^{\mathsf T} & 0
 \end{bmatrix}
 \begin{bmatrix}\bm C(\omega)\\\bm U(\omega)\\\Delta\mu(\omega)\end{bmatrix}
 &=\begin{bmatrix}\bm V(\omega)\\\bm0\\0\end{bmatrix},
 \label{eq:review-acks2-equations}
\end{align}
where the multiplier $\Delta\mu(\omega)$ is the change in chemical potential.
The last row enforces conservation of the total electron number.
The external-perturbation coefficients, density--potential overlaps, and electron-number integrals are defined by:
\begin{align}
    V_m(\omega)&=\int f_m(\bm r)\Delta v(\bm r,\omega)\,d\bm r,
    \nonumber\\
    O_{mn}&=\int f_m(\bm r)g_n(\bm r)\,d\bm r,
    \nonumber\\
    D_m&=\int f_m(\bm r)\,d\bm r.
    \label{eq:main-field-overlap-charge}
\end{align}
As in the density and potential expansions, setting $\omega=0$ recovers the static ACKS2 working equations.\cite{Verstraelen2014}

For the scalar-density formulation with local auxiliary potentials, the electronic-structure information enters through the following explicit hardness and implicit response matrices,
\begin{align}
    \eta^{\mathrm{exp}}_{mm'}(\omega)
    &=\iint f_m(\bm r)\eta^{\mathrm{exp}}(\bm r,\bm r',\omega)
          f_{m'}(\bm r')\,d\bm r\,d\bm r',
    \nonumber\\
    \chi^{\mathrm{imp}}_{nn'}(\omega)
    &=\iint g_n(\bm r)\chi^{\mathrm{imp}}(\bm r,\bm r',\omega)
          g_{n'}(\bm r')\,d\bm r\,d\bm r'.
    \label{eq:main-condensed-ingredients}
\end{align}
The two matrices describe complementary parts of the molecular response.
The implicit response, $\chi^{\mathrm{imp}}(\bm r,\bm r',\omega)$, describes how the auxiliary electronic system redistributes electron density when its one-electron potential is perturbed, whereas the explicit hardness, $\eta^{\mathrm{exp}}(\bm r,\bm r',\omega)$, describes how the explicitly treated interactions couple these density fluctuations.\cite{Cheng2022,Verstraelen2014}
The terms explicit and implicit refer to how the underlying energy functional depends on the density.
The explicit contribution is written directly as a functional of the density, whereas the implicit contribution depends on the density through the auxiliary electronic state generated by its potential.
Thus, identifying the implicit response with a noninteracting Kohn--Sham response is one particular realization of these working equations, rather than a requirement of ACKS2.
The density-matrix and nonlocal-operator generalizations used below preserve the matrix form of Eq.~\eqref{eq:review-acks2-equations}, with the corresponding hardness, response, and overlap matrices.

In this work, we use the bare independent-particle response constructed from the orbitals and orbital energies of the chosen HF or DFT reference as the implicit response.
It should be noted that this is the auxiliary response, not the fully interacting HF or hybrid response.
Its matrix elements at imaginary frequency $\omega=iu$, with $u\geq0$, are
\begin{align}
    \chi^{\mathrm{imp}}_{nn'}(iu)
    &=-4\sum_{i\in\mathrm{occ}}\sum_{a\in\mathrm{vir}}
      \frac{\Omega_{ia}}{\Omega_{ia}^2+u^2}
      \langle\phi_i|g_n|\phi_a\rangle
      \langle\phi_a|g_{n'}|\phi_i\rangle,
    \label{eq:main-implicit-response}
\end{align}
where $i,j$ denote occupied spatial orbitals and $a,b$ denote virtual spatial orbitals, with orbital energies $\epsilon_p$ and occupied--virtual energy differences $\Omega_{ia}=\epsilon_a-\epsilon_i$.
The diagonal matrix $\bm\Omega$ used below contains these energy differences.
The factor four includes the closed-shell spin sum.

For the special case of an adiabatic semilocal XC functional, the explicit hardness is
\begin{align}
    \eta^{\mathrm{exp}}(\bm r,\bm r')
    &=\frac{1}{|\bm r-\bm r'|}+f_{\mathrm{xc}}^{\mathrm{sl}}(\bm r,\bm r'),
    \label{eq:main-sem-local-kernel}
\end{align}
where $f_{\mathrm{xc}}^{\mathrm{sl}}(\bm r,\bm r')=\delta v_{\mathrm{xc}}^{\mathrm{sl}}(\bm r)/\delta\rho(\bm r')$ is the adiabatic semilocal XC kernel of time-dependent density-functional theory (TDDFT), evaluated at the reference density.
Here $v_{\mathrm{xc}}^{\mathrm{sl}}$ is the semilocal XC potential.
In this case, the explicit hardness is frequency independent.
In contrast, an exchange-containing reference requires the additional contribution derived in Sec.~\ref{subsec:main-exchange-response}.

Additionally, we choose biorthogonal density and potential basis functions, so that $\bm O=\bm I$, where $\bm I$ is the identity matrix.
It should be noted that the ACKS2$\omega$ working equations also apply when the two bases are not biorthogonal.
Within the electron-number-conserving response space and with the chosen biorthogonal bases, eliminating the auxiliary-potential coefficients gives the interacting density response:
\begin{align}
    \bm C(\omega)&=\bm\chi(\omega)\bm V(\omega),
    \nonumber\\
    \bm\chi(\omega)
    &=\left[\bm I-\bm\chi^{\mathrm{imp}}(\omega)
                   \bm\eta^{\mathrm{exp}}(\omega)\right]^{-1}
                   \bm\chi^{\mathrm{imp}}(\omega).
    \label{eq:main-acks2-response}
\end{align}
The matrix $\bm\chi$ represents the spatial response as
\begin{align}
    \chi(\bm r,\bm r',\omega)
    &\approx\sum_{mm'}f_m(\bm r)\chi_{mm'}(\omega)f_{m'}(\bm r').
    \label{eq:main-spatial-response}
\end{align}

\subsection{Response basis sets for ACKS2\texorpdfstring{$\omega$}{omega}}
\label{subsec:response-basis-sets}

In principle, the response basis sets for ACKS2 and ACKS2$\omega$ are not unique.\cite{Verstraelen2014,Cheng2022}
In this work, we use four response basis sets.
Set I is the original atomic multipole construction based on AIM partitioning from Ref.~\citenum{Verstraelen2014}.
Sets II and III add radial flexibility through local potentials or add density responses directly, respectively.
Finally, Set IV is the complete orbital-pair response basis within the chosen orbital basis and provides a full-space comparison.

It is important to distinguish the orbital basis used for HF or DFT from the functions used to describe an induced density in ACKS2.
The ACKS2 response basis determines which electronic rearrangements can be represented without changing the underlying orbital basis.
Table~\ref{tab:basis-terminology} summarizes the terminology used throughout the article.

\begin{table}[htbp]
\caption{Orbital-basis, ACKS2 density/potential-basis, and response-set terminology used throughout this work.}
\label{tab:basis-terminology}
\centering
\small
\renewcommand{\arraystretch}{1.15}
\begin{tabular}{>{\raggedright\arraybackslash}p{0.22\linewidth}>{\raggedright\arraybackslash}p{0.23\linewidth}>{\raggedright\arraybackslash}p{0.43\linewidth}}
\toprule
Term & Notation or name & Meaning in this work\\
\midrule
Orbital basis & $\{\varphi_\mu\}$, aug-cc-pVDZ & Atomic-orbital functions used to expand the molecular orbitals in every HF/DFT reference.\\
ACKS2 density basis & $\{f_m\}$, coefficients $C_m$ & Functions representing the induced electron density in ACKS2.\\
ACKS2 potential basis & $\{g_n\}$ or $\{\hat g_n\}$, coefficients $U_n$ & Local potentials of ACKS2 or, for the generalized constructions, nonlocal one-electron operators that generate the auxiliary response.\\
Set I: original response set from Ref.~\citenum{Verstraelen2014} & AIM multipoles and paired densities & Atom-centered multipole potentials and their biorthogonal density basis functions.\\
Set II: locally enlarged response set & Potential-first construction & The Set I potentials are supplemented by radial local potentials, and their static-response density partners satisfy the biorthogonality condition.\\
Set III: density-first response set & Density-first construction & The static Set I density-response span is supplemented by independent density-response components sampled at nonzero imaginary frequencies. Dual potential operators are constructed after the density basis is selected and may be nonlocal.\\
Set IV: complete orbital-pair response basis & $N_{\mathrm{ov}}=N_{\mathrm{occ}}N_{\mathrm{vir}}$ & One symmetric density-matrix variation per occupied--virtual pair, with a nonlocal dual potential operator.\\
Response-basis dimension & $N_r$ & Number of independent density-matrix variations that conserve electron number. This quantity is also called the response rank. An explicit Fukui pair is counted separately.\\
Full local candidate set & $N_{r,\mathrm{pool}}$ & All independent candidates in the specified Set II family. This set is not necessarily a complete response basis.\\
\bottomrule
\end{tabular}
\end{table}

\subsubsection{Set I: the original atomic response.}
To construct Set I, we use the following distributed multipole operators from the charge-flow analysis:\cite{Cheng2023}
\begin{align}
    g_{n(a,\ell,\kappa)}(\bm r)
    &=Q^a_{\ell\kappa}(\bm r)
      =w_a(\bm r)R_{\ell\kappa}(\bm r-\bm R^a),
    \nonumber\\
    \sum_a w_a(\bm r)&=1.
    \label{eq:main-atomic-potentials}
\end{align}
The weight function $0\leq w_a(\bm r)\leq1$ assigns a portion of the molecular density to atom $a$, whose nucleus is at $\bm R^a$.
The real regular solid harmonic $R_{\ell\kappa}$ specifies a multipole of order $\ell$ and component $\kappa$, with $R_{00}=1$.
The cutoff $\ell_{\max}$ sets the highest order included.
Monopole fluctuations describe changes in atomic populations and therefore the charge-flow part of the response, whereas dipoles and higher multipoles describe local polarization.

The corresponding density functions are then constructed from the chosen static independent-particle response to these potentials and supplemented by a normalized Fukui function:\cite{Verstraelen2014}
\begin{align}
    h_n(\bm r)
    &=\int\chi^{\mathrm{imp}}(\bm r,\bm r',0)g_n(\bm r')\,d\bm r',
    \nonumber\\
    f_{\mathrm F}(\bm r)
    &=\frac{1}{2}\left[|\phi_{\mathrm H}(\bm r)|^2
                       +|\phi_{\mathrm L}(\bm r)|^2\right],
    \nonumber\\
    \int f_{\mathrm F}(\bm r)\,d\bm r&=1.
    \label{eq:main-density-candidates}
\end{align}
Here $\phi_{\mathrm H}$ and $\phi_{\mathrm L}$ are the normalized highest occupied and lowest unoccupied molecular orbitals.
Each $h_n$ conserves electron number.
In contrast, the Fukui function represents a change in the total number of electrons, rather than polarization at fixed electron number.
Linear combinations of these candidates define $f_m$ so that $\int f_m(\bm r) g_n(\bm r)\,d\bm r=\delta_{mn}$ on the independent paired space.
In the original atomic multipole representation, $C_m=\int g_m(\bm r)\Delta\rho(\bm r)\,d\bm r$.
The electron-number integral $D_m$, defined in Eq.~\eqref{eq:main-field-overlap-charge}, is one for a monopole and zero otherwise.
The constraint therefore requires the sum of the atomic population changes to vanish.
Induced atomic charges have the opposite sign to these electron-population changes.\cite{Cheng2022,Verstraelen2014}

Increasing $\ell_{\max}$ adds angular multipoles.
The choice $\ell_{\max}=0$ includes population changes, $\ell_{\max}=1$ adds dipolar deformation, and higher orders allow more detailed angular rearrangements.
Before removing linear dependencies, there are $N_{\mathrm{pot}}=N_{\mathrm{atom}}(\ell_{\max}+1)^2$ Set I potentials.
The raw Set I list does not contain a separate constant function, but the monopoles contain the exact linear combination $\sum_a g_{a00}(\bm r)=\sum_a w_a(\bm r)=1$.
This constant combination has zero occupied--virtual coupling and therefore produces no fixed-electron-number response.
For the fixed-electron-number response we consequently use independent neutral linear combinations that span the same Set I response space after the single constant combination and any additional numerical dependencies are removed.
The dimension $N_r$ of this independent response space can therefore be smaller than the number of raw Set I functions.
When an explicit electron-number-changing degree of freedom is included, the constant potential is paired separately with the normalized Fukui function, giving $N_r+1$ density--potential pairs.
Further details of Set I are given in Section~S1.1 of the Supporting Information.

\subsubsection{Set II: local radial enrichment.}
Set I increases the angular flexibility of the response through higher multipoles, but each atomic multipole has a limited radial form.
Set II therefore adds radial flexibility, allowing the response within a given angular channel to differ close to and farther from the nucleus.
Starting from Set I, we add local radial potentials of the form
\begin{align}
 g^{\mathrm{rad}}_{a\ell\kappa\beta}(\bm r)
 &=Q^a_{\ell\kappa}(\bm r)
   \exp[-\beta|\bm r-\bm R^a|^2].
 \label{eq:main-radial-enrichment}
\end{align}
The different positive exponents control the radial extent while preserving the angular character of the underlying atomic multipole.
For $\ell=0$, these added functions can describe radial charge-shape fluctuations in addition to changes in net atomic population.
The complete Set II monopole space is therefore broader than interatomic charge flow alone.
The exponents and angular cutoff used in the calculations are specified in Sec.~\ref{sec:computational-details}.

The Set II generating set is the original Set I atomic-multipole potentials together with the added radial potentials in Eq.~\eqref{eq:main-radial-enrichment}.
For any potential block, let each column of $\bm G$ contain the occupied--virtual matrix elements $G_{ia,p}=\langle i|g_p|a\rangle$, so its rows correspond to occupied--virtual pairs $(i,a)$ and its columns correspond to potential functions.
The raw Set I transition block $\bm G_{\mathrm I}^{\mathrm{raw}}$ therefore has shape $N_{\mathrm{ov}}\times N_{\mathrm{pot}}$.
After removing the single constant monopole combination and any numerical dependencies, we denote the independent neutral Set I working block by $\bm G_0$, with shape $N_{\mathrm{ov}}\times N_{r,0}$.
Set I itself remains the original AIM generating set, while $\bm G_0$ is its independent fixed-electron-number representation.

With $\bm\Omega$ the diagonal matrix of positive orbital gaps $\epsilon_a-\epsilon_i$, the static independent-particle response defines the potential-space Gram matrix $\bm M^{(0)}=\bm G^{\mathsf T}\bm\Omega^{-1}\bm G=-\bm\chi^{\mathrm{imp}}(0)/4$.
Equivalently, $\bm M^{(0)}$ is the overlap matrix of the weighted transition vectors $\bm\Omega^{-1/2}\bm G$.
This metric compares potentials through the static occupied--virtual responses they generate, with low-energy transitions weighted more strongly, rather than through direct real-space overlap.

Let $\bm G_{\mathrm{rad}}$ contain the transition columns of all added radial candidates and let $\bm U_0$ span the weighted Set I space $\bm\Omega^{-1/2}\bm G_0$.
Projecting $\bm\Omega^{-1/2}\bm G_{\mathrm{rad}}$ onto the orthogonal complement of $\bm U_0$ removes the part of every radial candidate whose static independent-particle response is already represented by Set I.
A singular-value decomposition (SVD) of the remaining normalized columns removes null and near-linearly dependent combinations and produces an independent additional block $\bm G_{\mathrm{extra}}$.
The full neutral Set II working block is therefore $[\bm G_0,\bm G_{\mathrm{extra}}]$, while the physical Set II generating set remains the original Set I potentials plus the raw radial potentials.

Ordering is required only when a truncated Set II basis is used.
If $N_{\mathrm{extra}}=N_{r,\mathrm{pool}}-N_{r,0}$ and $k=\lceil\zeta N_{\mathrm{extra}}\rceil$, the selected working block is $\bm G_{\mathrm{II}}(\zeta)=[\bm G_0,\bm G_{\mathrm{extra}}^{(1:k)}]$ with dimension $N_r=N_{r,0}+k$.
Thus using $\zeta<1$ requires a nested ordering of the independent additional combinations, whereas the complete Set I block $\bm G_0$ is always included.
Two ordering schemes were examined for these independent additional combinations.
Static-potential ordering ranks the potential combinations directly, whereas response-informed ordering ranks them by how well they represent independent-particle response targets at three nonzero imaginary frequencies.
The corresponding response densities are
\begin{align}
 h_n(\bm r;iu_k)
 &=\int\chi^{\mathrm{imp}}(\bm r,\bm r',iu_k)g_n(\bm r')\,d\bm r',
 \label{eq:main-enrichment-responses}
\end{align}
with the three $u_k$ values specified in Sec.~\ref{sec:computational-details}.
These nonzero-frequency responses are used only to rank the already independent Set II potential combinations and are not themselves added as Set II density-basis functions.
No interacting-response, polarizability, or $C_6$ fit is used in this ordering.
After the potential set is selected, its paired density functions are constructed from the static independent-particle response.
To complete the density--potential representation, the normalized Fukui density $f_{\mathrm F}$ is paired separately with the constant potential $g_0=1$.
For each neutral potential $g_p$, we choose the constant-shift gauge $g'_p(\bm r)=g_p(\bm r)-a_p$, where $a_p=\int f_{\mathrm F}(\bm r)g_p(\bm r)\,d\bm r$, so that $\int f_{\mathrm F}(\bm r)g'_p(\bm r)\,d\bm r=0$.
This shift does not change any fixed-electron-number response because $\langle i|1|a\rangle=0$ for occupied and virtual orbitals, and it does not change the pairing with any neutral density because those densities integrate to zero.
The explicit constant/Fukui pair therefore completes the representation and fixes the gauge without adding an active fixed-$N$ response direction.
Throughout this work, the reported Set II rank $N_r$ and the matrices used for polarizabilities and ACFDT correlation energies contain only the neutral Set II subspace and exclude the constant/Fukui pair.
For $\zeta=1$, $\bm G_{\mathrm{II}}=[\bm G_0,\bm G_{\mathrm{extra}}]$, so all independent additional potentials are included and the ordering no longer changes the final Set II span.
Section~S2.1 of the Supporting Information gives the Set II construction and potential-ordering procedures in detail.
The values of $\zeta$ used in the calculations are given in Sec.~\ref{sec:computational-details}.

\subsubsection{Set III: density-first enrichment.}
Set III enlarges the response from the complementary direction by adding density-response functions directly rather than first enlarging the local potential space.
Using the independent neutral Set I block $\bm G_0\in\mathbb R^{N_{\mathrm{ov}}\times N_{r,0}}$ defined above, the static occupied--virtual density-response block is $\bm R(0)=\bm\Omega^{-1}\bm G_0$, which has the same shape and rank as $\bm G_0$.
Column normalization followed by SVD gives an orthonormal basis $\bm U_{\rho,0}$ for this complete static Set I density-response span.

At a nonzero imaginary frequency $u$, the corresponding independent-particle density-response shape is $\bm R(u)=\bm\Omega(\bm\Omega^2+u^2\bm I)^{-1}\bm G_0$, again with shape $N_{\mathrm{ov}}\times N_{r,0}$.
The common closed-shell factor $-2$ relating these response-shape blocks to the occupied--virtual blocks of the physical density-matrix response is omitted because it does not change the represented subspace or its SVD.
For the three nonzero frequencies specified in Sec.~\ref{sec:computational-details}, the blocks $\bm R(u_k)$ are concatenated and projected onto the ordinary orthogonal complement of $\bm U_{\rho,0}$ in occupied--virtual density space.
After column normalization, SVD removes null and near-linearly dependent residual components and gives an orthonormal additional density block $\bm U_{\rho,\mathrm{extra}}$.
The full available Set III density space is therefore $\bm U_\rho^{\mathrm{full}}=[\bm U_{\rho,0},\bm U_{\rho,\mathrm{extra}}]$.
For an equal-rank comparison with Set II, only the leading SVD-ordered additional density modes needed to reach the requested rank are included, while the complete static Set I density span is always preserved.
Unlike Set II, Set III has no separate $\zeta$ selection rule.

All columns of $\bm U_\rho$ are occupied--virtual density-matrix variations and therefore conserve electron number.
Set III consequently contains no explicit Fukui function.
After the density basis has been selected, compatible dual potential operators are constructed.
These operators are generally nonlocal, unlike the multiplicative local potentials of Set II.
Let the columns of $\bm U_\rho$ contain the selected orthonormal occupied--virtual density blocks.
Their paired potential blocks are
\begin{align}
 \bm U_g&=\frac{1}{2}\bm\Omega\bm U_\rho
       (\bm U_\rho^{\mathsf T}\bm\Omega\bm U_\rho)^{-1},
 \nonumber\\
 2\bm U_\rho^{\mathsf T}\bm U_g&=\bm I.
 \label{eq:main-density-first-pair}
\end{align}
The diagonal matrix $\bm\Omega$ contains the occupied--virtual orbital energy differences defined above.
Because $\bm U_\rho$ is orthonormal, $\bm U_\rho^{\mathsf T}\bm\Omega\bm U_\rho$ is positive definite on the selected density space, and the construction above gives $2\bm U_\rho^{\mathsf T}\bm U_g=\bm I$.
The factor two accounts for the occupied--virtual and transposed virtual--occupied blocks of a real symmetric density matrix.
Each density block and its transpose specify the associated density-matrix variation used in the exchange hardness.
For the nonlocal dual operators, Eq.~\eqref{eq:main-implicit-response} uses matrix elements of $\hat g_n$ in place of those of the local functions $g_n$.
Biorthogonality is therefore defined by a one-electron trace pairing rather than by the real-space product of two multiplicative scalar functions.
Section~S2.2 of the Supporting Information gives the matrix construction, SVD conditioning, and static-dual-potential derivation in detail.

A complete response basis must span all $N_{\mathrm{ov}}=N_{\mathrm{occ}}N_{\mathrm{vir}}$ independent occupied--virtual orbital-pair variations, where $N_{\mathrm{occ}}$ and $N_{\mathrm{vir}}$ are the numbers of occupied and virtual spatial orbitals.
However, equal nominal function counts do not prove completeness.
In particular, the finite frequency pool of Set III can saturate before reaching this limit, and the full local pool of Set II need not reach it either.

\subsubsection{Set IV: complete orbital-pair response basis.}
Set IV removes finite ACKS2 response-space truncation as a source of error within the chosen orbital basis.
It is constructed explicitly, rather than by extrapolating Set II or III, by including every independent occupied--virtual density-matrix variation.
For each occupied--virtual pair $(i,a)$ of the real orthonormal reference orbitals, we choose the symmetric density-matrix variation and its diagonal density function as
\begin{align}
 \hat F^{(ia)}&=|\phi_i\rangle\langle\phi_a|
                  +|\phi_a\rangle\langle\phi_i|,
 \nonumber\\
 f_{ia}(\bm r)&=\langle\bm r|\hat F^{(ia)}|\bm r\rangle
                =2\phi_i(\bm r)\phi_a(\bm r).
 \label{eq:main-complete-density-pair}
\end{align}
Orbital orthogonality gives $\int f_{ia}(\bm r)\,d\bm r=0$, so every variation conserves electron number and no Fukui function is needed.
The paired potential is the one-electron operator
\begin{align}
 \hat g_{ia}&=\frac{1}{2}\left(|\phi_i\rangle\langle\phi_a|
                            +|\phi_a\rangle\langle\phi_i|\right),
 \nonumber\\
 \operatorname{Tr}\!\left[\hat F^{(ia)}\hat g_{jb}\right]
    &=\delta_{ij}\delta_{ab}.
 \label{eq:main-complete-potential-pair}
\end{align}
Here the trace is over the one-electron orbital space.
For nonlocal potentials, this trace replaces the local overlap $O_{mn}=\int f_m(\bm r)g_n(\bm r)\,d\bm r$ defined in Eq.~\eqref{eq:main-field-overlap-charge}.
Defining the occupied--virtual blocks by $T_{jb,ia}=\langle\phi_j|\hat F^{(ia)}|\phi_b\rangle$ and $G_{jb,ia}=\langle\phi_j|\hat g_{ia}|\phi_b\rangle$ gives
\begin{align}
 \bm T&=\bm U_\rho=\bm I,
 \nonumber\\
 \bm G&=\bm U_g=\frac{1}{2}\bm I,
 \nonumber\\
 2\bm T^{\mathsf T}\bm G&=\bm I.
 \label{eq:main-complete-orbital-blocks}
\end{align}
This is the full-space specialization of Eq.~\eqref{eq:main-density-first-pair}.
The potential kernel is
\begin{align}
    g_{ia}(\bm r,\bm r')&=\frac{1}{2}\left[
 \phi_i(\bm r)\phi_a(\bm r')+\phi_a(\bm r)\phi_i(\bm r')\right].
 \label{eq:main-complete-potential-kernel}
\end{align}
It generally couples different positions and is not a multiplicative scalar potential.
The $N_{\mathrm{ov}}$ density-matrix variations are independent, even when their scalar diagonal functions $f_{ia}$ have linear dependencies.
Completeness here therefore refers to the specified density-matrix response space, not an assumed independent basis of scalar orbital products.

\subsection{Range-separated response and nonlocal exchange}
\label{subsec:main-exchange-response}

The HFsrPBE reference combines long-range HF exchange with the short-range PBE XC functional.\cite{Goll2005,Rebolini2013b}
Accordingly, the interaction is separated as $1/r=w^{\mathrm{lr}}(r)+w^{\mathrm{sr}}(r)$, where $w^{\mathrm{lr}}(r)=\operatorname{erf}(\mu r)/r$ and $w^{\mathrm{sr}}(r)=\operatorname{erfc}(\mu r)/r$.
The parameter $\mu$ controls the separation and has units of inverse length.
The same value is used in the reference and all its response ingredients.
The physical HFsrPBE response contains the full Hartree interaction, the short-range PBE kernel, and long-range exchange.
With the independent-particle auxiliary response chosen in Sec.~\ref{subsec:acks2-review}, these response interactions enter the explicit hardness.
Section~S3.1 of the Supporting Information describes the range-separated reference.

For reference, the density-basis Hartree and semilocal hardnesses are
\begin{align}
    \eta_{\mathrm H,mm'}
    &=\iint\frac{f_m(\bm r)f_{m'}(\bm r')}{|\bm r-\bm r'|}
               \,d\bm r\,d\bm r',
    \nonumber\\
    \eta_{\mathrm{xc},mm'}^{\mathrm{sl}}
    &=\iint f_m(\bm r)f_{\mathrm{xc}}^{\mathrm{sl}}(\bm r,\bm r')
                  f_{m'}(\bm r')\,d\bm r\,d\bm r'.
    \label{eq:main-hartree-sem-local-hardness}
\end{align}
The semilocal contribution is absent for HF.
For HFsrPBE, it contains short-range PBE XC, whereas for a global hybrid, it contains the corresponding semilocal part.

In contrast to the semilocal contribution, exact exchange depends on the full one-particle density matrix, rather than only on the electron density.
A scalar density fluctuation is therefore insufficient to specify how exchange changes when the electrons respond.
The corresponding density-matrix variation must also be specified.
In a real atomic-orbital basis $\{\varphi_\mu\}$, each density function therefore carries a real symmetric, spin-summed density-matrix variation $\bm F^{(m)}$:
\begin{align}
    f_m(\bm r)&=\sum_{\mu\nu}F^{(m)}_{\mu\nu}
                               \varphi_\mu(\bm r)\varphi_\nu(\bm r),
    \nonumber\\
    \Delta\bm P_{\mathrm{AO}}&=\sum_m C_m\bm F^{(m)}.
    \label{eq:main-density-matrix-functions}
\end{align}
For Sets I--III, these matrices are obtained with the response-generated density functions and undergo the same basis transformations.
For Set IV, they are the explicitly specified orbital-pair variations.
A fit to the scalar density alone would not specify them.
For real orbitals $\psi_p$, we use chemists' notation for the two-electron integrals:
\begin{align}
    (pq|rs)_w
    &=\iint\psi_p(\bm r)\psi_q(\bm r)w(|\bm r-\bm r'|)
           \psi_r(\bm r')\psi_s(\bm r')\,d\bm r\,d\bm r'.
    \label{eq:main-two-electron-integrals}
\end{align}
The exchange interaction $w_{\mathrm x}$ is $1/r$ for HF, $c_{\mathrm x}/r$ for a global hybrid with exact-exchange fraction $c_{\mathrm x}$, and $w^{\mathrm{lr}}$ for HFsrPBE.
The static exchange hardness is then
\begin{align}
    \eta_{\mathrm x,mm'}
    &=-\frac{1}{2}\sum_{\mu\nu\kappa\sigma}
       F^{(m)}_{\mu\nu}F^{(m')}_{\kappa\sigma}
       (\mu\kappa|\nu\sigma)_{w_{\mathrm x}}.
    \label{eq:main-static-exchange-hardness}
\end{align}
It describes the static curvature of the exchange energy with respect to the represented density-matrix fluctuations.
The prefactor follows from the spin-summed closed-shell convention.
The derivation is given in Section~S3.2 of the Supporting Information.

At nonzero frequency, exchange also changes the orbital mixing that accompanies the induced density.
This orbital relaxation is not contained in the static exchange hardness.
To describe it, we define the singlet orbital-response matrices as follows:\cite{Rebolini2013b}
\begin{align}
    \bm A&=\bm\Omega+2\bm J+2\bm F^{\mathrm{sl}}-\bm K_A,
    \nonumber\\
    \bm B&=2\bm J+2\bm F^{\mathrm{sl}}-\bm K_B.
    \label{eq:main-singlet-blocks}
\end{align}
Here $\Omega_{ia,jb}=\Omega_{ia}\delta_{ij}\delta_{ab}$, $J_{ia,jb}=(ia|jb)_{1/r}$, $(K_A)_{ia,jb}=(ij|ab)_{w_{\mathrm x}}$, and $(K_B)_{ia,jb}=(ib|aj)_{w_{\mathrm x}}$.
The semilocal matrix element is $F^{\mathrm{sl}}_{ia,jb}=\iint\phi_i(\bm r)\phi_a(\bm r) f_{\mathrm{xc}}^{\mathrm{sl}}(\bm r,\bm r') \phi_j(\bm r')\phi_b(\bm r')\,d\bm r\,d\bm r'$.
Thus $\bm A$ and $\bm B$ describe the interaction between the forward and backward orbital responses.
In their difference, $\bm A-\bm B=\bm\Omega-\bm K_A+\bm K_B$, the Hartree and adiabatic semilocal contributions cancel, but exchange generally does not.

To express this orbital response in density coordinates, we use the map defined by the same density matrices:
\begin{align}
    T_{ia,m}
    &=\left[\bm C_{\mathrm{MO}}^{\mathsf T}\bm S
              \bm F^{(m)}\bm S\bm C_{\mathrm{MO}}\right]_{ia},
    \nonumber\\
    \Delta P_{ia}^{\mathrm{MO}}&=\sum_m T_{ia,m}C_m.
    \label{eq:main-density-orbital-map}
\end{align}
The molecular orbitals satisfy $\phi_p=\sum_\mu(C_{\mathrm{MO}})_{\mu p}\varphi_\mu$.
Their atomic-basis overlap is $S_{\mu\nu}=\langle\varphi_\mu|\varphi_\nu\rangle$, with $\bm C_{\mathrm{MO}}^{\mathsf T}\bm S\bm C_{\mathrm{MO}}=\bm I$.
The matrix $\bm T$ has one row per occupied--virtual pair and one column per density function in the response basis.

Eliminating the difference-channel amplitudes and projecting the resulting correction onto the represented density-matrix variations gives the additional frequency-dependent exchange contribution to the hardness
\begin{align}
    \Delta\bm\eta_{\mathrm x}(iu)
    &=u^2\bm\Lambda_{\mathrm x},
    \nonumber\\
    \bm\Lambda_{\mathrm x}
    &=\bm T^{\mathsf T}
      \left[(\bm A-\bm B)^{-1}-\bm\Omega^{-1}\right]\bm T.
    \label{eq:main-exchange-hardness}
\end{align}
The inverse is taken for a stable reference with a nonsingular $\bm A-\bm B$.
The frequency-independent matrix $\bm\Lambda_{\mathrm x}$ is constructed once for a fixed reference and density basis.
Combining the Hartree, semilocal, and exchange contributions then gives
\begin{align}
    \bm\eta^{\mathrm{exp}}(iu)
    &=\bm\eta_{\mathrm H}+\bm\eta_{\mathrm{xc}}^{\mathrm{sl}}
      +\bm\eta_{\mathrm x}+\Delta\bm\eta_{\mathrm x}(iu).
    \label{eq:main-explicit-hardness}
\end{align}
Using this hardness in Eq.~\eqref{eq:main-acks2-response} gives the exchange-corrected ACKS2$\omega$ response.
The added term vanishes at $u=0$ and for an adiabatic semilocal response, for which $\bm A-\bm B=\bm\Omega$.
It therefore leaves the static response at a fixed basis unchanged.
Its frequency dependence arises from the orbital elimination, not from an assumed memory dependence of the exchange functional.
The orbital elimination and projection onto the density basis are derived in Sections~S3.3 and S3.4 of the Supporting Information, respectively.

\subsection{Molecular polarizability and charge fluctuations}
\label{subsec:main-response-properties}

The molecular dipole polarizability provides a direct connection between the distributed electronic response and a molecular observable.
For the original atomic population monopoles, we use the terms charge fluctuations and charge flow equivalently.
The atomic multipole representation also contains local-polarization contributions and the coupling between charge flow and local polarization.
With the electron-number convention, the induced dipole is $\Delta\mu_i=-\int r_i\Delta\rho(\bm r)\,d\bm r$.
For a uniform field $\bm{\mathcal E}$, the perturbation is $\Delta v=\bm r\cdot\bm{\mathcal E}$, giving $V_m=\sum_i L_{mi}\mathcal E_i$, where $L_{mi}=\int f_m(\bm r)r_i\,d\bm r$, and hence $\bm\alpha=-\bm L^{\mathsf T}\bm\chi\bm L$.
The same minus sign is used in the distributed polarizability with the tensor labels of the charge-flow analysis:\cite{Cheng2022,Cheng2023}
\begin{align}
    \alpha^{ab}_{\ell\kappa,\ell'\kappa'}(\omega)
    &=-\iint Q^a_{\ell\kappa}(\bm r)
        \chi(\bm r,\bm r',\omega)
        Q^b_{\ell'\kappa'}(\bm r')\,d\bm r\,d\bm r'.
    \label{eq:main-distributed-polarizability}
\end{align}
Here $a,b$ again label atoms.
The label $0$ denotes a monopole and $i,j\in\{x,y,z\}$ denote Cartesian dipoles, with $Q^a_0=w_a$ and $Q^a_i=w_a(r_i-R_i^a)$.
The blocks $\alpha^{ab}_{00}$, $\alpha^{ab}_{0j}$, $\alpha^{ab}_{i0}$, and $\alpha^{ab}_{ij}$ describe charge flow, the two mixed charge--dipole responses, and dipole--dipole response, respectively.
In a general density--potential basis, the potential-operator response is $\bm\chi'=\bm O^{\mathsf T}\bm\chi\bm O$.
The distributed polarizabilities are directly the negatives of the entries of $\bm\chi$ only when the paired basis functions carry the corresponding atomic multipole labels.

The molecular dipole polarizability can then be reconstructed from these distributed contributions as\cite{Cheng2023}
\begin{align}
    \alpha_{ij}(\omega)
    &=\sum_{ab}\left[
      R_i^aR_j^b\alpha^{ab}_{00}(\omega)
      +R_i^a\alpha^{ab}_{0j}(\omega)
      +R_j^b\alpha^{ab}_{i0}(\omega)
      +\alpha^{ab}_{ij}(\omega)\right].
    \label{eq:main-molecular-reconstruction}
\end{align}
The charge-flow contribution is the first term of this sum.
For two molecules $A$ and $B$, the isotropic $C_6$ dispersion coefficient is
\begin{align}
    C_6^{AB}
    &=\frac{3}{\pi}\int_0^\infty
      \bar\alpha^A(iu)\bar\alpha^B(iu)\,du,
    \nonumber\\
    \bar\alpha(iu)&=\frac{1}{3}\operatorname{Tr}\bm\alpha(iu).
    \label{eq:review-casimir-polder}
\end{align}

\subsection{Long-range correlation energy}
\label{subsec:acfdt-correlation}

Long-range correlation can be viewed as the energetic consequence of coupling electronic fluctuations throughout the system.
The frequency-dependent response developed above describes these fluctuations, while ACFDT connects their coupling to the correlation energy.
For reference, ACFDT expresses the full-range correlation energy as\cite{Langreth1975,Gunnarsson1976}
\begin{align}
    E_{\mathrm c}
    &=-\frac{1}{2\pi}\int_0^1d\lambda\int_0^\infty du
      \iint\frac{\chi_\lambda(\bm r,\bm r',iu)
                    -\chi_{\mathrm s}(\bm r,\bm r',iu)}
                   {|\bm r-\bm r'|}\,d\bm r\,d\bm r'.
    \label{eq:main-acfdt-fullrange}
\end{align}
Here $\lambda\in[0,1]$ scales the Coulomb interaction while the external potential is adjusted to keep the ground-state density fixed, and $u\geq0$ is the imaginary-frequency integration variable.
The interacting density response is $\chi_\lambda$, and $\chi_{\mathrm s}=\chi_{\lambda=0}$ is the noninteracting Kohn--Sham response for that density.
Their difference removes the fluctuations already present in the noninteracting reference.

In this work, we apply the corresponding response-difference contraction to a long-range, frozen-reference approximation, rather than evaluating the exact density-preserving path in Eq.~\eqref{eq:main-acfdt-fullrange}.
In this approximation, $\lambda$ scales only the long-range electron--electron interaction.
For a post-SCF correction to HFsrPBE, the complementary short-range correlation is already included in the reference energy.
We therefore couple the fluctuations through $w^{\mathrm{lr}}(r)=\operatorname{erf}(\mu r)/r$, with the same $\mu$ as the reference.
The resulting energy is $E_{\mathrm{HFsrPBE}}+E_{\mathrm c}^{\mathrm{lr}}$.

The long-range interaction between two density functions is the ACKS2 Hartree hardness
\begin{align}
    \eta^{\mathrm{lr}}_{\mathrm H,mn}
    &=\iint f_m(\bm r)
      \frac{\operatorname{erf}(\mu|\bm r-\bm r'|)}{|\bm r-\bm r'|}
      f_n(\bm r')\,d\bm r\,d\bm r'.
    \label{eq:main-longrange-hardness}
\end{align}
It is the Coulomb interaction between the electronic rearrangements represented by $f_m$ and $f_n$, restricted to its long-range part.
It should be noted that this matrix is distinct from the full Hartree and XC hardness in Eqs.~\eqref{eq:main-hartree-sem-local-hardness} and~\eqref{eq:main-explicit-hardness}, which enter the physical HFsrPBE polarizability.

To express the energy in ACKS2 parameters, we evaluate all response and hardness matrices in the same fixed, biorthogonal density--potential basis, restricted to variations that conserve electron number.
The HFsrPBE reference, density and potential functions, and bare independent-particle auxiliary response $\bm\chi^{\mathrm{imp}}(iu)$ from Eq.~\eqref{eq:main-implicit-response} are held fixed while the long-range interaction is introduced.
Section~S4.1 of the Supporting Information derives the density-basis contraction used for the correlation energy.

\subsubsection{Long-range response including exchange.}
We include exchange by applying the correction of Sec.~\ref{subsec:main-exchange-response} at each coupling strength.
The explicit long-range hardness is
\begin{align}
    \bm\eta_{\lambda}^{\mathrm{exp,lr}}(iu)
    &=\lambda\bm\eta_{\mathrm H}^{\mathrm{lr}}
      +\lambda\bm\eta_{\mathrm x}^{\mathrm{lr}}
      +\Delta\bm\eta_{\mathrm x,\lambda}^{\mathrm{lr}}(iu),
    \nonumber\\
    \Delta\bm\eta_{\mathrm x,\lambda}^{\mathrm{lr}}(iu)
    &=u^2\bm\Lambda_{\mathrm x}^{\mathrm{lr}}(\lambda).
    \label{eq:main-longrange-exchange-hardness}
\end{align}
Here $\bm\eta_{\mathrm x}^{\mathrm{lr}}$ is the static exchange hardness defined in Eq.~\eqref{eq:main-static-exchange-hardness}, with $w_{\mathrm x}=w^{\mathrm{lr}}$.
Its linear scaling follows from scaling that interaction while keeping the density functions fixed.
The matrix $\bm\Lambda_{\mathrm x}^{\mathrm{lr}}(\lambda)$ supplies the additional exchange response at nonzero frequency.
Using the same density-to-orbital transformation $\bm T$ and matrix of orbital energy differences $\bm\Omega$, this parameter is obtained as
\begin{align}
    \bm\Lambda_{\mathrm x}^{\mathrm{lr}}(\lambda)
    &=\bm T^{\mathsf T}
      \left[
      \left(\bm\Omega-\lambda\bm K_A^{\mathrm{lr}}
                         +\lambda\bm K_B^{\mathrm{lr}}\right)^{-1}
      -\bm\Omega^{-1}\right]\bm T.
    \label{eq:main-longrange-exchange-parameter}
\end{align}
The long-range exchange elements are $(K_A^{\mathrm{lr}})_{ia,jb}=(ij|ab)_{w^{\mathrm{lr}}}$ and $(K_B^{\mathrm{lr}})_{ia,jb}=(ib|aj)_{w^{\mathrm{lr}}}$, with $i,j$ occupied and $a,b$ virtual orbitals and the integral convention of Eq.~\eqref{eq:main-two-electron-integrals}.
The matrix being inverted must be nonsingular at every coupling strength used in the integration.
Equation~\eqref{eq:main-longrange-exchange-parameter} constructs the condensed exchange parameter for use in the density response.
It should be noted that the interaction must be scaled before this inverse is evaluated.
Consequently, $\bm\Lambda_{\mathrm x}^{\mathrm{lr}}(\lambda)$ is not generally $\lambda\bm\Lambda_{\mathrm x}^{\mathrm{lr}}(1)$.
Parameters at the required coupling strengths are therefore needed, not only the full-coupling correction.

The coupled density response entering the energy is now specified by
\begin{align}
    \bm\chi_\lambda(iu)
    &=\left[\bm I-\bm\chi^{\mathrm{imp}}(iu)
                      \bm\eta_{\lambda}^{\mathrm{exp,lr}}(iu)\right]^{-1}
                      \bm\chi^{\mathrm{imp}}(iu).
    \label{eq:main-longrange-coupled-response}
\end{align}
At $\lambda=0$, both exchange contributions and the Hartree contribution vanish, giving $\bm\chi_0=\bm\chi^{\mathrm{imp}}$.
In contrast, at $u=0$, only the additional frequency-dependent exchange term vanishes.
The static exchange hardness remains for $\lambda>0$.
This response includes long-range Hartree and exchange interactions.
It does not include the full Hartree and short-range PBE response kernels used for the physical HFsrPBE polarizability.
The short-range PBE treatment enters the reference, rather than being added again to this long-range response kernel.\cite{Toulouse2010}

\subsubsection{Correlation energy with ACKS2$\omega$.}
Using this coupled response, we express the long-range correlation energy as
\begin{align}
    E_{\mathrm c}^{\mathrm{lr}}[\bm\chi_\lambda]
    &=-\frac{1}{2\pi}\int_0^1d\lambda\int_0^\infty du\,
      \operatorname{Tr}\!\left[
        \left(\bm\chi_\lambda(iu)-\bm\chi^{\mathrm{imp}}(iu)\right)
        \bm\eta_{\mathrm H}^{\mathrm{lr}}\right].
    \label{eq:main-acfdt-condensed-energy}
\end{align}
The trace sums over the paired density indices $m,n$.
Exchange enters through $\bm\chi_\lambda$, whereas the change in fluctuations is weighted by the long-range Coulomb interaction, $\bm\eta_{\mathrm H}^{\mathrm{lr}}$.
Replacing this energy weight by the total response hardness would define a different approximation.
An exchange-containing response with this direct Coulomb contraction is the RPAx-I choice.\cite{Angyan2011}
Section~S4.2 of the Supporting Information explains the distinction between exchange in the response and Coulomb weighting of the energy.
Equations~\eqref{eq:main-longrange-exchange-hardness}--\eqref{eq:main-acfdt-condensed-energy} specify the ACKS2$\omega$ approximation to RPAx-I using the projected exchange correction.
In particular, the added exchange hardness is projected in the fixed density basis, whereas the independent-particle auxiliary response is projected before inversion.
Projection and inversion do not generally commute in a finite response basis (Section~S3 of the Supporting Information).
Once $\bm\chi^{\mathrm{imp}}(iu)$, $\bm\eta_{\mathrm H}^{\mathrm{lr}}$, $\bm\eta_{\mathrm x}^{\mathrm{lr}}$, and $\bm\Lambda_{\mathrm x}^{\mathrm{lr}}(\lambda)$ have been prepared, the energy integration uses only ACKS2 matrices.

\subsubsection{Direct RPA as a special case.}
In direct RPA, the induced fluctuations interact only through the long-range Hartree term, $\bm\eta_{\lambda}^{\mathrm{exp,lr}}=\lambda\bm\eta_{\mathrm H}^{\mathrm{lr}}$.
Consequently,
\begin{align}
    \bm\chi_{\lambda,\mathrm d}(iu)
    &=\left[\bm I-\lambda\bm\chi^{\mathrm{imp}}(iu)
                      \bm\eta_{\mathrm H}^{\mathrm{lr}}\right]^{-1}
                      \bm\chi^{\mathrm{imp}}(iu),
    \label{eq:main-drpa-response}\\
    E_{\mathrm c}^{\mathrm{lr,dRPA-I}}
    &=-\frac{1}{2\pi}\int_0^1d\lambda\int_0^\infty du\,
      \operatorname{Tr}\!\left[
        \left(\bm\chi_{\lambda,\mathrm d}(iu)-\bm\chi^{\mathrm{imp}}(iu)\right)
        \bm\eta_{\mathrm H}^{\mathrm{lr}}\right].
    \label{eq:main-drpa-energy}
\end{align}
The label dRPA-I specifies a Hartree-only response kernel and the direct, non-antisymmetrized interaction in the energy expression.\cite{Angyan2011}
This special case sets both $\bm\eta_{\mathrm x}^{\mathrm{lr}}$ and $\bm\Lambda_{\mathrm x}^{\mathrm{lr}}(\lambda)$ to zero, even when HFsrPBE orbitals were used to construct the reference parameters.
This is a choice of response approximation, not a restriction imposed by using ACKS2 parameters.
The physical short-range PBE response kernel is not added to this direct-RPA hardness either.

For this special case, the coupling-strength integration can be performed analytically, giving an energy that uses only the two ACKS2 ingredients $\bm\chi^{\mathrm{imp}}(iu)$ and $\bm\eta_{\mathrm H}^{\mathrm{lr}}$:
\begin{align}
    E_{\mathrm c}^{\mathrm{lr,dRPA-I}}
    &=\frac{1}{2\pi}\int_0^\infty du\,
      \left\{\ln\det\!\left[\bm I-
         \bm\chi^{\mathrm{imp}}(iu)\bm\eta_{\mathrm H}^{\mathrm{lr}}\right]
       +\operatorname{Tr}\!\left[
         \bm\chi^{\mathrm{imp}}(iu)\bm\eta_{\mathrm H}^{\mathrm{lr}}\right]
      \right\}.
    \label{eq:main-drpa-logdet}
\end{align}
All matrix operations are performed in the ACKS2 density-response basis.
The derivation of this logarithmic expression can be found in Section~S4.3 of the Supporting Information.
It provides an independent check of the numerical coupling-strength integration for dRPA-I.

\subsubsection{RPA with exchange (RPAx-I).}
In contrast to dRPA-I, the exchange-containing RPAx-I choice includes both the static exchange hardness and the additional frequency-dependent exchange term.
In the same electron-number-conserving response basis, Eqs.~\eqref{eq:main-longrange-coupled-response} and~\eqref{eq:main-acfdt-condensed-energy} become
\begin{align}
    \bm\chi_{\lambda,\mathrm x}(iu)
    &=\left\{
      \left[\bm\chi^{\mathrm{imp}}(iu)\right]^{-1}
      -\lambda\left(\bm\eta_{\mathrm H}^{\mathrm{lr}}
                    +\bm\eta_{\mathrm x}^{\mathrm{lr}}\right)
      -u^2\bm\Lambda_{\mathrm x}^{\mathrm{lr}}(\lambda)
      \right\}^{-1},
    \nonumber\\
    E_{\mathrm c}^{\mathrm{lr,RPAx-I}}
    &=-\frac{1}{2\pi}\int_0^1d\lambda\int_0^\infty du\,
      \operatorname{Tr}\!\left[
        \left(\bm\chi_{\lambda,\mathrm x}(iu)-\bm\chi^{\mathrm{imp}}(iu)\right)
        \bm\eta_{\mathrm H}^{\mathrm{lr}}\right].
    \label{eq:main-rpax-response-energy}
\end{align}
Thus, dRPA-I and RPAx-I use the same Hartree-only energy contraction but different coupled responses.
The suffix I denotes this direct, non-antisymmetrized contraction, not omission of exchange from the response.\cite{Angyan2011}
With a reduced ACKS2 response basis, these expressions define the projected approximation to orbital RPAx-I.
Equivalence to the orbital reference requires the complete occupied--virtual response space.
Setting both $\bm\eta_{\mathrm x}^{\mathrm{lr}}$ and $\bm\Lambda_{\mathrm x}^{\mathrm{lr}}(\lambda)$ to zero recovers Eqs.~\eqref{eq:main-drpa-response} and~\eqref{eq:main-drpa-energy}.
Keeping the static exchange hardness while dropping only $\bm\Lambda_{\mathrm x}^{\mathrm{lr}}(\lambda)$ instead defines the static-exchange-only approximation to RPAx-I, used here as a diagnostic.

To validate Set IV, we use an independent orbital-space ACFDT calculation with the same frozen HFsrPBE orbitals and long-range interaction.
The working equations for this orbital-space calculation can be found in Section~S4.4 of the Supporting Information.

\subsubsection{Charge-fluctuation contributions.}
The trace form also permits the long-range correlation energy to be analyzed in terms of the local electronic fluctuations represented by the atomic moments.
For the original atomic population monopoles, charge fluctuations and charge flow refer to the same monopole response, consistent with the distributed-polarizability analysis.
For a specified atomic multipole representation, let $\mathcal I_0$ contain the monopole indices and $\mathcal I_{>0}$ contain all higher multipoles.
With $\Delta\bm\chi_\lambda=\bm\chi_\lambda-\bm\chi^{\mathrm{imp}}$, we define
\begin{align}
    E_{\mathrm c}^{XY}
    &=-\frac{1}{2\pi}\int_0^1d\lambda\int_0^\infty du\,
      \sum_{m\in\mathcal I_X}\sum_{n\in\mathcal I_Y}
      \Delta\chi_{\lambda,mn}(iu)\eta^{\mathrm{lr}}_{\mathrm H,nm},
    \qquad X,Y\in\{0,>0\}.
    \label{eq:main-correlation-multipole-parts}
\end{align}
The total is $E_{\mathrm c}^{00}+E_{\mathrm c}^{0,>0} +E_{\mathrm c}^{>0,0}+E_{\mathrm c}^{>0,>0}$.
Formally, these are the monopole--monopole, mixed monopole--multipole, and multipole--multipole contributions extracted from the same coupled response, with both orders included for the mixed term.
For the original population monopoles, the first two groups have the usual charge--charge and mixed charge--multipole interpretation.
In Set II, however, the $\ell=0$ group also contains radial charge-shape fluctuations, so the complete monopole block is broader than net interatomic charge flow alone.

For Set II, the chemically labelled moment frame used for the energy decomposition is deliberately distinct from the independent working basis used to solve the response.
The working block $\bm G_{\mathrm{II}}=[\bm G_0,\bm G_{\mathrm{extra}}]$ contains linear combinations generated by conditioning, projection, and SVD, so its individual columns are not assigned monopole, dipole, or higher-multipole labels.
Instead, let $g_\alpha^{\mathrm{raw}}$ denote one operator in the chemically labelled frame formed from the original AIM multipoles together with the full radial multipole pool, and let $\bm G_\alpha^{\mathrm{raw}}$ be its occupied--virtual transition vector.
We normalize each individual moment operator by its static independent-particle metric norm,
\begin{align}
 s_\alpha
 &=\left[(\bm G_\alpha^{\mathrm{raw}})^{\mathsf T}
          \bm\Omega^{-1}\bm G_\alpha^{\mathrm{raw}}\right]^{1/2},
 \nonumber\\
 \widetilde g_\alpha&=g_\alpha^{\mathrm{raw}}/s_\alpha,
 \nonumber\\
 (G_{\mathrm{mom}})_{ia,\alpha}
 &=\langle\phi_i|\widetilde g_\alpha|\phi_a\rangle.
 \label{eq:main-moment-frame-normalization}
\end{align}
Because all orbital gaps are positive, $s_\alpha=0$ if and only if that individual operator has zero occupied--virtual coupling, and such an operator cannot be normalized.
This condition does not require the complete moment frame to be linearly independent.
In particular, the original atomic monopoles can each have nonzero $s_\alpha$ while their sum generates the constant potential and hence a collective null combination.
The resulting $\bm G_{\mathrm{mom}}$ therefore preserves the chemical labels of the original and radial moments even though it may be redundant.
The vector $\bm q$ collects the values of these normalized local moments for a density represented in the independent Set II response coordinates.
For the density map $\bm T$ defined above,
\begin{align}
 \bm q&=\bm M_{\mathrm{mom}}\bm C,
 \nonumber\\
 \bm M_{\mathrm{mom}}&=2\bm G_{\mathrm{mom}}^{\mathsf T}\bm T,
 \nonumber\\
 \bm N_{\mathrm{mom}}&=\bm M_{\mathrm{mom}}^+,
 \nonumber\\
 \bm N_{\mathrm{mom}}\bm M_{\mathrm{mom}}&=\bm I.
 \label{eq:main-moment-frame-maps}
\end{align}
The moment map $\bm M_{\mathrm{mom}}$ is generally rectangular because the chemically labelled moment frame can contain more operators than the independent Set II response rank.
The superscript $+$ denotes its Moore--Penrose pseudoinverse.
The identity $\bm N_{\mathrm{mom}}\bm M_{\mathrm{mom}}=\bm I$ requires $\bm M_{\mathrm{mom}}$ to have full column rank, meaning that the local moment frame spans every density variation in the selected electron-number-conserving response basis even though the moment frame itself may contain collective linear dependencies.
The pseudoinverse then supplies the minimum-norm inverse mapping used to express the Hartree interaction in this redundant chemical frame.
The response and minimum-norm Hartree matrices in this local-moment representation are
\begin{align}
 \bm\chi_\lambda^{\mathrm{mom}}
 &=\bm M_{\mathrm{mom}}\bm\chi_\lambda\bm M_{\mathrm{mom}}^{\mathsf T},
 \nonumber\\
 \bm\eta_{\mathrm H}^{\mathrm{lr,mom}}
 &=\bm N_{\mathrm{mom}}^{\mathsf T}\bm\eta_{\mathrm H}^{\mathrm{lr}}
   \bm N_{\mathrm{mom}}.
 \label{eq:main-moment-frame-response-hardness}
\end{align}
The same map is applied to the independent-particle auxiliary response.
Equation~\eqref{eq:main-correlation-multipole-parts} is applied to the matrices expressed in terms of local atomic moments, whose indices identify monopole or higher-multipole contributions.
Since $\bm N_{\mathrm{mom}}\bm M_{\mathrm{mom}}=\bm I$, the full trace equals the original ACKS2 trace.
Equivalently, if $\bm S_X$ is diagonal with ones on the moment indices in group $X$ and zeros elsewhere, and $\bm P_X=\bm N_{\mathrm{mom}}\bm S_X\bm M_{\mathrm{mom}}$, each contribution can be evaluated in the smaller ACKS2 response basis as
\begin{align}
 \sum_{\alpha\in\mathcal I_X}\sum_{\beta\in\mathcal I_Y}
 \Delta\chi^{\mathrm{mom}}_{\lambda,\alpha\beta}
 \eta^{\mathrm{lr,mom}}_{\mathrm H,\beta\alpha}
 &=\operatorname{Tr}\!\left[
 \bm P_X\Delta\bm\chi_\lambda\bm P_Y^{\mathsf T}
 \bm\eta_{\mathrm H}^{\mathrm{lr}}\right].
 \label{eq:main-moment-frame-block-trace}
\end{align}

Finally, to resolve the dipolar part, we split $\mathcal I_{>0}$ into the dipole indices $\mathcal I_1$ and the remaining indices $\mathcal I_{\geq2}$.
Applying the same contraction in Eq.~\eqref{eq:main-correlation-multipole-parts} to these subsets gives
\begin{align}
    E_{\mathrm c}^{>0,>0}
    &=E_{\mathrm c}^{11}+E_{\mathrm c}^{\mathrm{rem}},
    \nonumber\\
    E_{\mathrm c}^{\mathrm{rem}}
    &=E_{\mathrm c}^{1,\geq2}+E_{\mathrm c}^{\geq2,1}
      +E_{\mathrm c}^{\geq2,\geq2}.
    \label{eq:main-dipole-remainder-split}
\end{align}
The dipole indices include the original and radially enriched $\ell=1$ moments.
The local moment operators and the pseudoinverse remain fixed.
This decomposition separates contributions to the same coupled response, rather than introducing a separate dipole-only response calculation.
In particular, the remainder includes both dipole--higher-multipole orders, not only higher--higher interactions.
 \section{Computational details}
\label{sec:computational-details}
\subsection{Electronic-structure calculations}

We studied the molecular polarizabilities and dispersion coefficients of the TS27 set using HF, PBE, B3LYP, and HFsrPBE with the aug-cc-pVDZ basis.
The optimized molecular geometries are from the earlier ACKS2$\omega$ study.\cite{Cheng2022,Toulouse2013}
For HFsrPBE, the range-separation parameter was $\mu=0.4\;\mathrm{bohr}^{-1}$ in the ground-state calculation, the short-range XC kernel, and the long-range exchange integrals.

For the TS27 calculations, the HF, PBE, and B3LYP input orbitals and static polarizabilities were obtained with Gaussian 16, Revision C.01.\cite{Gaussian16}
The HFsrPBE ground-state calculations used HORTON\cite{HORTON211} with our in-house Python package \texttt{LRC-CD}.
Independent HFsrPBE reference polarizabilities at 13 imaginary frequencies were calculated with TDDFT in Dalton2020.1 (2022).\cite{Aidas2014,Dalton2020}

The ACKS2$\omega$ calculations and matching full orbital-response references used HORTON and \texttt{LRC-CD} for all 27 molecules and all four methods, as well as for the 12-molecule equal-response-basis-size comparison of Sets II and III.
The orbital response was evaluated using time-dependent Hartree--Fock (TDHF) for HF and TDDFT for PBE, B3LYP, and HFsrPBE.
The semilocal XC contributions to the explicit hardness for PBE, B3LYP, and HFsrPBE were evaluated by finite differences following Ref.~\citenum{Verstraelen2014}, using a step of $10^{-4}$ in this work.
Further numerical protocols are given in Section~S5 of the Supporting Information.
The SCF consistency criteria for the 12-molecule and energy comparisons, molecular-grid settings for the energy calculations, and independent response and energy checks are described in Sections~S5.1--S5.3, respectively.

Table~\ref{tab:reported-calculation-settings} summarizes the response calculations.
Within each comparison, all response sets used identical electronic references and numerical settings.

\begin{table}[htbp]
\centering
\small
\begin{tabular}{>{\raggedright\arraybackslash}p{0.28\linewidth}>{\raggedright\arraybackslash}p{0.25\linewidth}>{\raggedright\arraybackslash}p{0.16\linewidth}>{\raggedright\arraybackslash}p{0.19\linewidth}}
\toprule
Reported comparison & Systems and response sets & Molecular grid & Frequency and $C_6$ sampling\\
\midrule
HFsrPBE versus TDDFT (Dalton) & TS27, Set I, $\ell_{\max}=0,\ldots,4$ & \texttt{medium}, seed 0 & 13 frequencies, 12 nonzero quadrature nodes\\
Local enrichment & TS27 with HF, PBE, B3LYP, and HFsrPBE using Sets I/II & \texttt{veryfine}, seed 0 & 91 frequencies, 48-point $C_6$ quadrature\\
Exchange comparison & \ce{H2O}/HF, \ce{CO}/HF, and \ce{CO}/B3LYP using Set II at $\zeta=0.80$ & \texttt{veryfine}, seed 0 & 91 frequencies, 48-point $C_6$ quadrature\\
Density-first comparison & 12 molecules with HF, B3LYP, and HFsrPBE using Sets I--IV & \texttt{veryfine}, seed 0 & 187 frequencies, 96-point $C_6$ quadrature\\
\bottomrule
\end{tabular}
\caption{Electronic-reference scope, response sets, HORTON molecular grids, and frequency/$C_6$ sampling used for each reported molecular-response comparison.}
\label{tab:reported-calculation-settings}
\end{table}

\subsection{Response sets and frequency integration}

The density and potential functions of ACKS2$\omega$ were constructed as described in Sec.~\ref{subsec:response-basis-sets}, using atomic weights from minimal basis iterative stockholder (MBIS) partitioning.\cite{Verstraelen2016}
For every molecule, the local construction included atomic multipoles through $\ell_{\max}=4$ and added the three Set II radial factors with $\beta=0.15$, $0.6$, and $2.4\;\mathrm{bohr}^{-2}$.
The raw Set I potentials were transformed to their independent fixed-electron-number span after removal of the constant monopole combination.
The added radial candidates were then projected onto the orthogonal complement of this Set I span in the static independent-particle metric $\mathbf G^{\mathsf T}\bm\Omega^{-1}\mathbf G$.
Column-normalized SVD with a relative cutoff of $10^{-8}$ removed remaining null and near-linearly dependent radial combinations while preserving the complete independent Set I span.
For truncated Set II representations, both the static-potential and response-informed ordering schemes described in Sec.~\ref{subsec:response-basis-sets} were tested.
Across the selection tests, the fractions $\zeta=0.50,0.65,0.80,0.90,0.95,1.00$ were examined, and the response-informed ordering at $\zeta=0.80$ was selected for the subsequent truncated Set II calculations.
The full local pool includes all independent additional combinations, so its final span is independent of the ordering.
The explicit constant-potential/Fukui pair completes the Set II density--potential representation but is excluded from all reported fixed-electron-number response ranks and from the polarizability and ACFDT response matrices.
Section~S6 and Tables~S1 and S2 of the Supporting Information document the selection procedure and its response errors.

The exchange comparison in Table~\ref{tab:local-exchange-ablation} uses the same Set II response-informed potential ordering at $\zeta=0.80$, with radial multipoles through order 4 for all three systems.
The response-basis dimensions are 159 for \ce{H2O}/HF and 169 for \ce{CO}/HF and \ce{CO}/B3LYP.
The functions are held fixed when including or omitting the frequency-dependent exchange term.

Set III started from the complete static Set I density-response span $\bm\Omega^{-1}\bm G_0$ and added independent density-response components sampled at $u=0.649114$, $5.957643$, and $32.239080$ a.u.
The three nonzero-frequency response blocks were projected onto the orthogonal complement of the static density span, normalized, and conditioned by SVD with relative cutoff $10^{-10}$.
For intermediate Set III ranks, the static density block was always included and the leading SVD-ordered additional frequency-sampled modes were added until the requested rank was reached.
No separate $\zeta$ selection was used for Set III.
In the small-molecule comparisons, the full available Set III dimensions were 180 for \ce{H2O} and 196 for \ce{CO} and \ce{N2}, compared with complete-response dimensions of 180 and 273, respectively.
All Set III density responses integrate to zero, so this construction conserves electron number and uses no explicit Fukui function.

The 91 frequencies used to evaluate the Set II frequency-dependent dipole polarizabilities comprise the original 13 frequencies, the six additional values $u=0.02,0.1,0.5,2,10,100$, and the nodes of 24- and 48-point quadratures.
These six values test the response outside the three nonzero frequencies used only for response-informed ordering of the Set II potential candidates.
Homomolecular $C_6$ coefficients were calculated from Eq.~\eqref{eq:review-casimir-polder} with $A=B$.
Gauss--Legendre quadrature on $x\in[-1,1]$ used the transformation $u=0.3(1+x)/(1-x)$ with $du/dx=0.6/(1-x)^2$.
The local-enrichment results use 48 points.
The independent Dalton comparison uses the original 12-node transformed quadrature.

\subsection{Comparison of Sets II and III at equal basis size}

Sets II and III were compared for \ce{H2O}, CO, \ce{N2}, \ce{CCl4}, \ce{CO2}, \ce{SO2}, \ce{Cl2}, \ce{HBr}, \ce{CS2}, COS, \ce{CH3CHO}, and \ce{N2O}, using HF, B3LYP, and HFsrPBE.
For this comparison, the maximum absolute occupied--virtual Fock element was required to be below $10^{-6}$ hartree, evaluated with the HORTON Hamiltonian used for the response.
The maximum absolute difference between the Fock matrix in the molecular-orbital basis and the diagonal matrix of orbital energies was required to satisfy the same threshold.
All 36 electronic references used in this comparison satisfied both criteria.
The SCF convergence settings are given in Section~S5.1 of the Supporting Information.

Because Sets II and III can provide different maximum numbers of independent functions, we compared them using the smaller of the Set II size at $\zeta=0.80$ and the available Set III size.
Additional equal-size comparisons included the full Set III pool, with Set IV as the complete-response reference.
The reported dimension excludes the separate Fukui pair in Sets I and II.
The 187 evaluation frequencies add the 96-point quadrature nodes to the earlier 91-frequency set.
The reported $C_6$ coefficients use 96 points, with relative changes below $10^{-8}$ compared with 48 points.
Section~S5.3 of the Supporting Information describes the independent orbital-response checks.

\subsection{Long-range correlation energies}

Long-range correlation energies were calculated for \ce{H2O}, \ce{CO}, and \ce{N2} using HFsrPBE/aug-cc-pVDZ with $\mu=0.4\;\mathrm{bohr}^{-1}$.
The electronic references satisfied the same SCF consistency criteria as the 12-molecule response comparison.
The calculations used the HORTON \texttt{veryfine} molecular grid with seed zero.
We compared Set I, the 80\% Set II selection, the full local pool, and Set III with the complete orbital-pair response basis defined in Sec.~\ref{subsec:response-basis-sets}.
The complete basis has $\bm T=\bm I$ and $\bm G=\bm I/2$, with 180 density--potential pairs for \ce{H2O} and 273 for \ce{CO} and \ce{N2}.
Set III used 159 and 180 functions for \ce{H2O} and 169 and 196 for the diatomics.
Set II was also evaluated with 196 functions for \ce{CO} and \ce{N2}, providing equal-size comparisons in addition to its full local basis of 199 functions.
The three exchange treatments used identical response functions within each comparison.

Both Hartree and exchange parameters use $\operatorname{erf}(\mu r_{12})/r_{12}$ for the long-range energy path.
The short-range PBE response kernel used for physical HFsrPBE polarizabilities is excluded from this path.
Coupling strength enters the exchange parameter before the difference-channel matrix is inverted.
After preparing the parameters, the ACKS2 energy is evaluated entirely from condensed response and hardness matrices.
The independent orbital-space calculation constructs its response from long-range two-electron integrals evaluated directly with the same orbitals.

The final quadrature uses 24 Gauss--Legendre points for coupling strength and 96 for imaginary frequency, with the frequency transformation given above.
The quadratures were varied independently over 8, 16, and 24 coupling points and 24, 48, and 96 frequency points.
An absolute energy comparison tolerance of $10^{-9}$ hartree was used.
Further numerical checks are described in Section~S5.3 of the Supporting Information.

The energy contributions for \ce{H2O}, \ce{CO}, and \ce{N2} were obtained in the local moment representation defined in Sec.~\ref{subsec:acfdt-correlation}.
It contains the original AIM operators and the full radial potential pool, with each individual operator normalized by its nonzero static independent-particle metric norm.
The complete moment frame may remain collectively linearly dependent, including the constant linear combination of the original atomic monopoles.
The rectangular moment map was required to have full column rank over the selected Set II response space and was inverted with a Moore--Penrose inverse at relative cutoff $10^{-10}$.
Added radial $\ell=0$ functions describe charge-shape changes and were included with the original atomic-population moments in the monopole group.
The $\ell=1$ moments define the dipole group.
The monopole--monopole, combined mixed, dipole--dipole, and remaining contributions were extracted from the same coupled response and sum to its full-pool Set II energy.
The dependence on molecular grid and on the definition of the moment operators was examined separately for \ce{H2O}.

\subsection{Measures of molecular-response agreement}

We report the relative errors below as numerical percentages.
For the isotropic polarizability, the signed percentage error and its maximum absolute value are:
\begin{align}
 \delta_\alpha(u)&=\left[
 \frac{\bar\alpha_{\mathrm{model}}(iu)}{\bar\alpha_{\mathrm{ref}}(iu)}-1\right] \times 100\%,
 \nonumber\\
 E_{\max}&=\max_{u\in\mathcal U}|\delta_\alpha(u)|.
 \label{eq:validation-error}
\end{align}
Here $\bar\alpha_{\mathrm{model}}$ and $\bar\alpha_{\mathrm{ref}}$ are the isotropic polarizabilities from ACKS2$\omega$ and the corresponding orbital-response reference, respectively.
The $C_6$ error is defined in the same way.
The set $\mathcal U$ contains the frequencies of the stated comparison.
To include all field directions, not only the isotropic average, we compare the polarizability tensors $\bm\alpha_{\mathrm{model}}$ and $\bm\alpha_{\mathrm{ref}}$ through:
\begin{align}
 \bm M_\alpha(u)&=\bm\alpha_{\mathrm{ref}}(iu)^{-1/2}
  [\bm\alpha_{\mathrm{model}}(iu)-\bm\alpha_{\mathrm{ref}}(iu)]
  \bm\alpha_{\mathrm{ref}}(iu)^{-1/2},
 \label{eq:relative-polarizability-tensor}\\
 E_{\mathrm{dir}}(u)&=\max_j|\lambda_j[\bm M_\alpha(u)]| \times 100\%.
 \label{eq:directional-polarizability-error}
\end{align}
The reference tensor is positive definite, and $\bm\alpha_{\mathrm{ref}}^{-1/2}$ is its symmetric inverse square root.
The dimensionless matrix $\bm M_\alpha$ measures the tensor difference relative to the reference response.
Its eigenvalues are denoted by $\lambda_j$, with $j$ running over the three spatial components.
The quantity $E_{\mathrm{dir}}$ is the largest absolute relative polarizability error over all directions of the applied field.
The two 1\% criteria used in the response-set comparison concern $\max_u E_{\mathrm{dir}}$ and $|\delta_{C_6}|$.

The reported scalar anisotropy is:
\begin{align}
 \gamma_\alpha(iu)&=\sqrt{\frac32}
   \|\bm\alpha(iu)-\bar\alpha(iu)\bm I\|_{\mathrm F}.
 \label{eq:polarizability-anisotropy}
\end{align}
The subscript F denotes the Frobenius norm.
Its relative error is $\delta_\gamma=(\gamma_{\alpha,\mathrm{model}}/ \gamma_{\alpha,\mathrm{ref}}-1)\times 100\%$.
It is left undefined when the reference anisotropy is smaller than $\bar\alpha_{\mathrm{ref}} \times 10^{-8}$.
For the real symmetric tensors here, $\gamma_\alpha=(3/\sqrt2)\|\vec\alpha\|$ in the recoupled notation of Cheng and Verstraelen.\cite{Cheng2023}
The anisotropies and errors reported here use this normalization.
Maximum errors are taken over the discrete frequency sets specified in Table~\ref{tab:reported-calculation-settings}.
 \section{Results}
\label{sec:numerical-validation}
We compare molecular polarizabilities, dispersion coefficients, and correlation energies with the corresponding orbital-response results in the same orbital basis.
All ACKS2$\omega$ calculations below use the independent-particle auxiliary response specified in Sec.~\ref{subsec:acks2-review}.

\subsection{Exchange and frequency-dependent molecular polarization}
\label{subsec:frequency-validation}

In the present construction, the additional exchange hardness vanishes at zero frequency, but changes the induced electron redistribution at nonzero frequency.
The distinction matters for dispersion because $C_6$ depends on the polarizability over the imaginary-frequency range, not only its static value.

\begin{figure}[!t]
\centering
\includegraphics[width=3.33in]{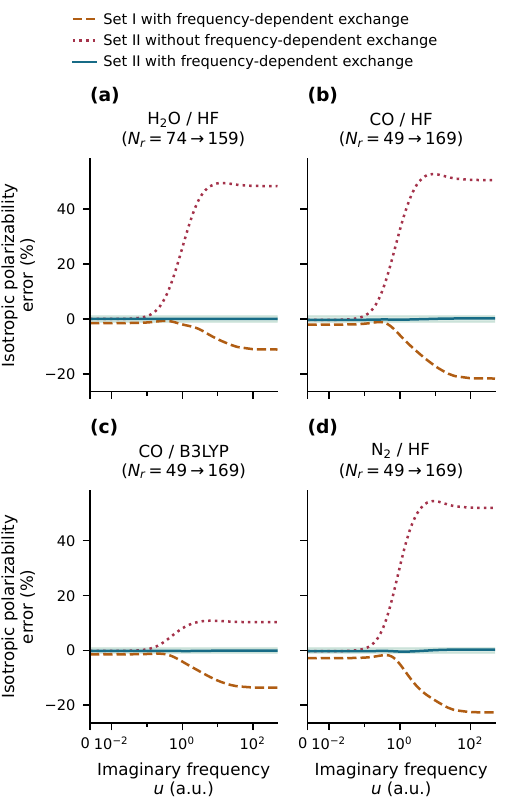}
\caption{Frequency-dependent polarizability errors using the same local selection rule for all systems shown.
Set I is the original response basis.
Set II includes the local radial enrichment at $\zeta=0.80$.
All curves show isotropic polarizability errors relative to the matched full response.
The comparison with and without frequency-dependent exchange changes only that contribution to the hardness.
Lines connect the calculated values at the sampled imaginary frequencies, and the green shaded region marks relative errors between $-1\%$ and $+1\%$.}
\label{fig:exchange-basis-errors}
\end{figure}

Table~\ref{tab:local-exchange-ablation} isolates this effect in \ce{H2O} and \ce{CO}.
It uses the same Set II selection at $\zeta=0.80$ as Fig.~\ref{fig:exchange-basis-errors}, with response-basis dimensions of 159 for \ce{H2O}/HF and 169 for both CO references (Table~S2 of the Supporting Information).
The density and potential functions, bare independent-particle auxiliary response, and static hardness are held fixed while only the additional frequency-dependent exchange contribution is included or omitted.
With HF, omitting the term overestimates the polarizability at the illustrated nonzero frequency by about 49--51\%, even in the enlarged local representation.
The corresponding $C_6$ coefficients are too large by about 27--31\%.
Including the term reduces the largest absolute $C_6$ error among these three systems to 0.52\%.
B3LYP shows the same pattern, with a smaller but still appreciable effect.

\begin{table}[htbp]
\centering
\small
\caption{Percentage errors in the isotropic dipole polarizability at $u=32.239080\,\mathrm{a.u.}$ and in $C_6$ for Set II at $\zeta=0.80$, relative to the matched full response, comparing calculations with and without the frequency-dependent exchange contribution while keeping the static exchange hardness in both.}
\label{tab:local-exchange-ablation}
\setlength{\tabcolsep}{4pt}
\begin{tabular}{lrrrrr}
\toprule
& & \multicolumn{2}{c}{$\delta_\alpha(u=32.239080\,\mathrm{a.u.})$ (\%)}
 & \multicolumn{2}{c}{$\delta_{C_6}$ (\%)}\\
\cmidrule(lr){3-4}\cmidrule(lr){5-6}
System & $N_r$ & \makecell{Without frequency-\\dependent exchange} & \makecell{With frequency-\\dependent exchange} & \makecell{Without frequency-\\dependent exchange} & \makecell{With frequency-\\dependent exchange}\\
\midrule
\ce{H2O}/HF & 159 & $+48.96$ & $+0.01$ & $+26.94$ & $-0.02$\\
\ce{CO}/HF & 169 & $+51.27$ & $+0.15$ & $+31.12$ & $-0.52$\\
\ce{CO}/B3LYP & 169 & $+10.41$ & $-0.04$ & $+7.28$ & $-0.32$\\
\bottomrule
\end{tabular}
\end{table}

Figure~\ref{fig:exchange-basis-errors} shows that the exchange correction reduces the error, but finite-basis errors remain with Set I.
The enlarged basis and the exchange correction address these two distinct sources of error.

\subsection{Range-separated polarization and dispersion}
\label{subsec:srpbe-validation}

HFsrPBE provides the reference relevant to the long-range correlation formulation.
We first use the original response set (Set I) and compare its molecular response with independent TDDFT calculations performed with Dalton.
For all 27 molecules in TS27 at $\ell_{\max}=4$, including the additional exchange term reduces the absolute $C_6$ error and both the largest directional and isotropic polarizability errors.
The median signed $C_6$ error changes from $+24.19\%$ to $-2.38\%$ (Table~\ref{tab:srpbe-ts27-production}).
Thus the excessive dispersion strength obtained with static exchange alone is substantially reduced.
In Table~\ref{tab:srpbe-ts27-production}, the directional error is the maximum over the 13 sampled frequencies.
The ``both'' criterion requires both the directional and $C_6$ errors to be below 1\%.

\begin{table}[htbp]
\centering
\small
\setlength{\tabcolsep}{5pt}
\begin{tabular}{lrrrr}
\toprule
Response treatment & Median directional error (\%) & Median $\delta_{C_6}$ (\%) &
 $|\delta_{C_6}|<1\%$ & Both criteria $<1\%$\\
\midrule
\makecell{Without frequency-\\dependent exchange} & $49.08\%$ & $+24.19\%$ & $0/27$ & $0/27$\\
\makecell{With frequency-\\dependent exchange} & $13.59\%$ & $-2.38\%$ & $9/27$ & $0/27$\\
\bottomrule
\end{tabular}
\caption{HFsrPBE/aug-cc-pVDZ Set I response at $\ell_{\max}=4$ and $\mu=0.4\;\mathrm{bohr}^{-1}$ relative to independent Dalton TDDFT over 13 imaginary frequencies, reporting median directional and $C_6$ errors and the numbers of molecules meeting the 1\% $C_6$ and joint criteria.}
\label{tab:srpbe-ts27-production}
\end{table}

\begin{figure}[!t]
\centering
\includegraphics[width=\linewidth]{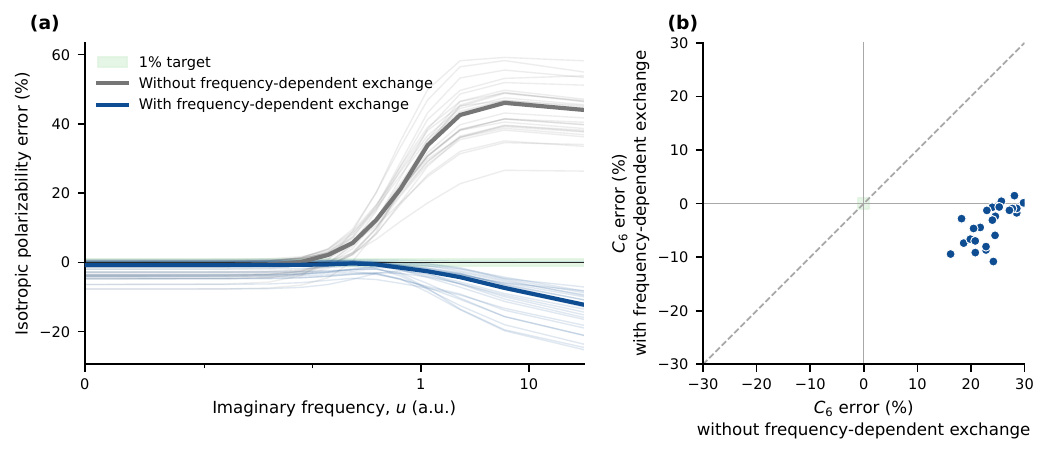}
\caption{HFsrPBE polarization and dispersion with the original response set.
(a) Signed isotropic polarizability errors at 13 imaginary frequencies.
Thin lines represent molecules and thick lines their medians.
(b) Paired homomolecular $C_6$ errors with and without the additional exchange term.
All 27 molecules are included.
The functional, aug-cc-pVDZ orbital basis, and $\mu=0.4\;\mathrm{bohr}^{-1}$ match the TDDFT reference computed with Dalton.
The green regions mark errors between $-1\%$ and $+1\%$.
The diagonal in (b) denotes equal signed errors.}
\label{fig:srpbe-ts27-production}
\end{figure}

Figure~\ref{fig:srpbe-ts27-production} shows the molecule-resolved frequency-dependent polarizability errors and paired $C_6$ errors underlying these aggregate statistics.

None of these original-set calculations meets both 1\% criteria against the independent TDDFT reference.
This cross-code comparison shows the improvement from adding the exchange term, but does not isolate the source of the remaining discrepancy.
Numerical orbitals, dipole representations, and response implementations can also differ.
To isolate response-set enlargement, we next compare Sets I and II using identical HFsrPBE orbitals and orbital energies.

\begin{figure}[!t]
\centering
\includegraphics[width=4.85in]{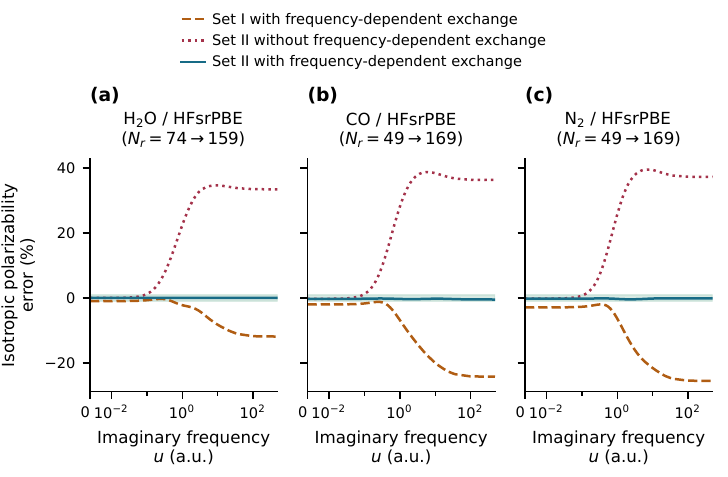}
\caption{Effects of enriching Set I to Set II and including the additional exchange hardness on HFsrPBE frequency-dependent polarization.
The Set I and Set II calculations use the same HFsrPBE orbitals and orbital energies.
Errors are relative to the corresponding full orbital response.
The green shaded region marks relative polarizability errors between $-1\%$ and $+1\%$.}
\label{fig:srpbe-exchange-basis}
\end{figure}

For \ce{H2O}, \ce{CO}, and \ce{N2}, the same local selection rule fixed using the HF and B3LYP cases gives the results in Table~\ref{tab:srpbe-transfer}.
Their largest isotropic errors decrease from 11.87\%, 24.22\%, and 25.53\% with Set I to 0.02\%, 0.48\%, and 0.43\% with Set II, respectively.
The small $C_6$ errors show that the improved frequency-dependent polarizability also yields more accurate dispersion coefficients.
All errors in Table~\ref{tab:srpbe-transfer} are percentages, and the maxima include all 91 sampled frequencies.
Here $N_r$ excludes the separate Fukui pair, and $\delta_\gamma$ is the relative anisotropy error.
Figure~\ref{fig:srpbe-exchange-basis} illustrates this comparison over the sampled imaginary frequencies.

\begin{table}[htbp]
\centering
\small
\begin{tabular}{llrrrr}
\toprule
System & $N_r/N_{\mathrm{ov}}$ & $E_{\max}$ (\%) &
 $\max_u E_{\mathrm{dir}}$ (\%) & $\delta_{C_6}$ (\%) & $\max_u|\delta_\gamma|$ (\%)\\
\midrule
\ce{H2O} & $159/180$ & $0.02$ & $0.03$ & $-0.02$ & $1.78$\\
\ce{CO} & $169/273$ & $0.48$ & $0.86$ & $-0.46$ & $3.60$\\
\ce{N2} & $169/273$ & $0.43$ & $0.60$ & $-0.42$ & $2.21$\\
\bottomrule
\end{tabular}
\caption{Percentage errors for HFsrPBE/aug-cc-pVDZ Set II at $\zeta=0.80$ and $\mu=0.4\;\mathrm{bohr}^{-1}$ relative to the matched full response over 91 imaginary frequencies, where $N_r$ excludes the Fukui pair and $\delta_\gamma$ is the relative anisotropy error.}
\label{tab:srpbe-transfer}
\end{table}

At the same enlarged ranks, omitting the frequency-dependent term gives $C_6$ errors of $+22.36\%$, $+26.23\%$, and $+24.08\%$ for \ce{H2O}, \ce{CO}, and \ce{N2}.
Including it gives the much smaller values in Table~\ref{tab:srpbe-transfer}.
Thus, with the auxiliary response used here, HFsrPBE, like HF and B3LYP, requires the frequency-dependent exchange contribution even with an enlarged response basis.

\subsection{Local response flexibility across molecular environments}
\label{subsec:full-basis-transferability}

Set II allows additional radial deformations of each atom's contribution to the electron density.
We applied the same construction to all TS27 molecules with HF, PBE, B3LYP, and HFsrPBE.
The orbital basis, angular cutoff, radial exponents, and selection rule were unchanged between molecules.
PBE provides a semilocal comparison, for which the additional frequency-dependent exchange term is zero.
The question is whether the extra local flexibility remains useful across the different molecular references, not whether the molecule-specific response parameters are transferable to a new molecule.

Table~\ref{tab:full-basis-transferability} shows a substantial improvement in both polarization and dispersion.
For Set II, we keep all independent Set I functions and add 80\% of the additional independent local candidates ($\zeta=0.80$).
With this selection, the median directional polarizability error falls from 13.40\% to 0.30\% for HF, from 7.90\% to 0.18\% for B3LYP, and from 14.25\% to 0.42\% for HFsrPBE.
The corresponding median absolute $C_6$ errors are 0.19\%, 0.08\%, and 0.09\%.
Thus local radial flexibility is useful for all three exchange-containing references, rather than only for the molecules used to fix the selection rule.
The full local pool includes every independent local candidate.

\begin{table}[htbp]
\centering
\small
\setlength{\tabcolsep}{5pt}
\begin{tabular}{lrrrrr}
\toprule
& \multicolumn{3}{c}{Molecules (of 27) with both errors below 1\%}
 & \multicolumn{2}{c}{Median error (\%), Set II (80\%)}\\
\cmidrule(lr){2-4}\cmidrule(lr){5-6}
Method & Set I & Set II (80\%) & Full local pool & Directional & $|\delta_{C_6}|$\\
\midrule
HF & 1 & 18 & 18 & 0.30 & 0.19\\
PBE & 12 & 24 & 27 & 0.03 & 0.02\\
B3LYP & 1 & 18 & 18 & 0.18 & 0.08\\
HFsrPBE & 1 & 18 & 18 & 0.42 & 0.09\\
\bottomrule
\end{tabular}
\caption{TS27 counts of 27 molecules satisfying both $\max_uE_{\mathrm{dir}}<1\%$ and $|\delta_{C_6}|<1\%$ for Set I, Set II at $\zeta=0.80$, and the full Set II local pool, together with the median percentage errors for the $\zeta=0.80$ selection.}
\label{tab:full-basis-transferability}
\end{table}

\begin{figure}[htbp]
\centering
\includegraphics[width=\linewidth]{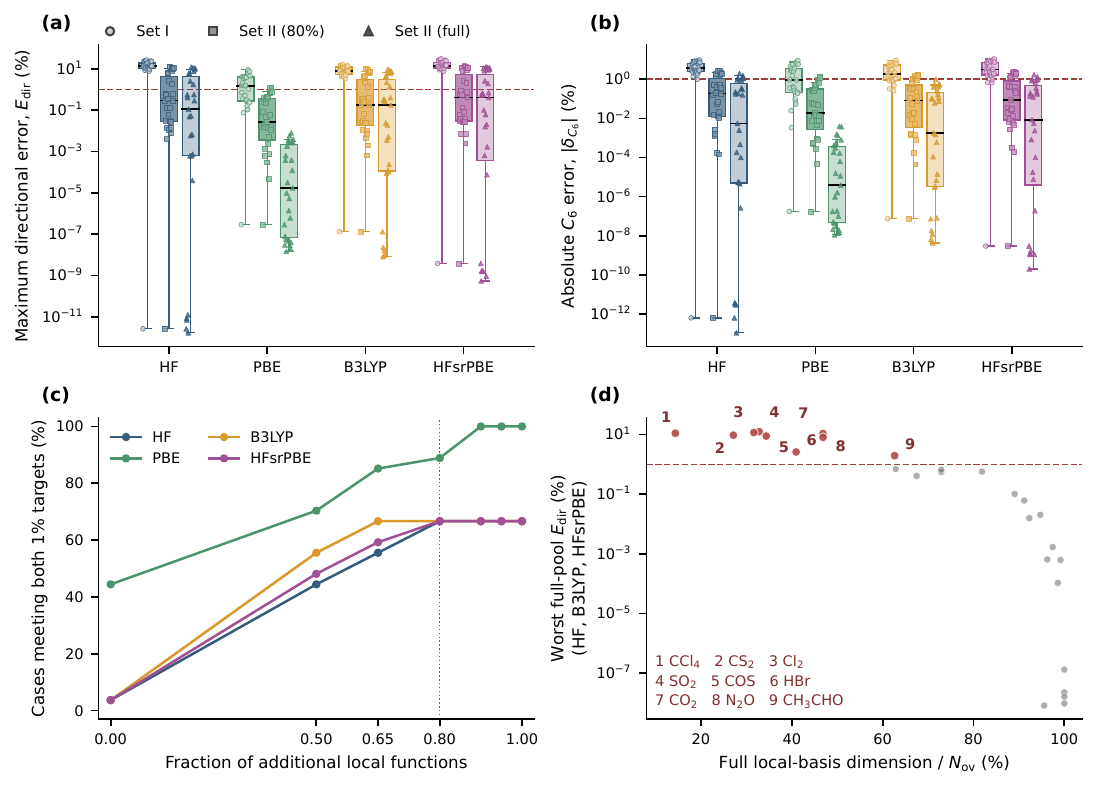}
\caption{Effect and limits of local radial enrichment.
The three selections are Set I, Set II with 80\% of the additional independent local candidates, and Set II with all such candidates.
Panels (a) and (b) compare directional polarization and $C_6$ errors.
Panel (c) shows how many molecules satisfy both 1\% criteria as the response set grows.
Panel (d) compares the dimension of the full local basis with $N_{\mathrm{ov}}$ and identifies molecules whose directional errors still exceed 1\% with at least one exchange-containing method.
Red dashed horizontal lines in (a), (b), and (d) mark the 1\% error threshold.
This is a comparison of response representations in a fixed orbital basis.}
\label{fig:full-basis-transferability}
\end{figure}

The absolute $C_6$ error decreases in every case where the original response space could be enlarged.
Directional errors decrease in 102 of those 104 method--molecule cases.
The exceptions are \ce{CCl4}/B3LYP and \ce{Cl2}/B3LYP.
The \ce{H2} response space was already saturated for all four references.
The summary therefore includes both the median errors and the number of molecules meeting the accuracy criteria.
Per-case response ranks, errors, pass flags, numerical tests, and warnings for all 108 TS27 method--molecule cases are given in Section~S7 and Tables~S3--S5 of the Supporting Information.
Figure~\ref{fig:full-basis-transferability} shows the molecule-resolved errors and the limits of the full local pool.

With every independent local candidate included, Set II meets both criteria for all PBE molecules but only 18 of 27 molecules for each of HF, B3LYP, and HFsrPBE.
The same nine molecules fall outside these criteria, namely \ce{CCl4}, \ce{CO2}, \ce{SO2}, \ce{Cl2}, \ce{HBr}, \ce{CS2}, \ce{COS}, \ce{CH3CHO}, and \ce{N2O}.
They include several different chemical environments.
The present aggregate comparison does not assign the residual error to a particular bond or functional group.
Their local pools cover only 14.35--62.63\% of the reference occupied--virtual spaces, and their worst directional errors among the three exchange-containing methods lie between 1.97 and 12.61\%.

The remaining errors show the limits of the chosen radial functions, not of local potentials in general.
We next compare them with density-first functions at equal basis size.

\subsection{Density-first enrichment and the complete-response limit}
\label{subsec:density-first-audit}

Set III adds response densities directly, with compatible dual potentials, instead of first enlarging the local perturbing-potential space.
We compare the two constructions for the nine molecules outside the earlier local-pool criteria, together with the three small-molecule test cases \ce{H2O}, CO, and \ce{N2}.
The comparisons use stationary HF, B3LYP, and HFsrPBE references.
At each matched rank, the density and potential functions differ between Sets II and III.
Only the dimension and electronic reference are shared.
Table~\ref{tab:density-first-matched} summarizes the target-molecule comparison, and Section~S8 of the Supporting Information gives the full response comparisons.

\begin{table}[htbp]
\centering\small
\begin{tabular}{lrrrrrr}
\toprule
& \multicolumn{2}{c}{HF} & \multicolumn{2}{c}{B3LYP} & \multicolumn{2}{c}{HFsrPBE}\\
Molecule & II & III & II & III & II & III\\
\midrule
\ce{CCl4} & 10.78 & 3.18 & 7.62 & 2.59 & 11.18 & 3.06 \\
\ce{CO2} & 6.61 & 1.98 & 5.01 & 0.74 & 7.55 & 1.07 \\
\ce{SO2} & 12.89 & 2.97 & 9.95 & 1.52 & 13.38 & 1.84 \\
\ce{Cl2} & 11.32 & 5.36 & 8.54 & 4.14 & 12.33 & 4.94 \\
\ce{HBr} & 2.72 & 0.94 & 2.28 & 0.81 & 2.74 & 1.07 \\
\ce{CS2} & 10.17 & 3.19 & 7.65 & 2.79 & 10.65 & 3.47 \\
\ce{COS} & 8.24 & 2.40 & 6.03 & 1.70 & 9.21 & 2.15 \\
\ce{CH3CHO} & 1.82 & 0.79 & 1.41 & 0.21 & 2.07 & 0.34 \\
\ce{N2O} & 6.05 & 2.18 & 4.99 & 1.04 & 7.35 & 1.39 \\
\bottomrule
\end{tabular}
\caption{Maximum directional polarizability errors (percent) for the nine local-pool target molecules comparing Sets II and III at equal response-basis dimension, where the matched rank is the selected Set II $\zeta=0.80$ rank or the smaller available Set III rank and Tables~S6--S8 of the Supporting Information give the full-pool dimensions, all matched-rank errors, and absolute response quantities.}
\label{tab:density-first-matched}
\end{table}
 
Set III reduced the maximum directional polarizability error in all 27 target molecule/method pairs (Table~\ref{tab:density-first-matched}).
For \ce{CCl4}/HFsrPBE, the error decreased from 11.18\% to 3.06\% at rank 424.
For \ce{CO2}/HFsrPBE, it decreased from 7.55\% to 1.07\% at rank 254.
The latter comparison is resolved into fixed principal-axis components in Figure~S1 of the Supporting Information.
The improvement did not extend uniformly to the three small-molecule test cases.
CO/HF and \ce{N2} with all three methods had larger directional errors with Set III at the selected matched rank.

Smaller directional errors did not ensure better dispersion coefficients.
The absolute $C_6$ error decreased in only five of the 27 target pairs.
These cases are \ce{SO2} and \ce{CH3CHO} for HF and HFsrPBE, and \ce{CH3CHO} for B3LYP.
For \ce{CCl4}/HFsrPBE, the signed $C_6$ error instead changed from $-1.94\%$ to $-4.05\%$, despite the smaller directional error.
At the selected matched ranks, Set III satisfied both 1\% criteria for \ce{HBr} and \ce{CH3CHO} with HF, for \ce{CO2}, \ce{HBr}, and \ce{CH3CHO} with B3LYP, and for \ce{CH3CHO} with HFsrPBE.
Set II satisfied the joint criterion for none of the nine targets.
Table~S6 of the Supporting Information lists the full Set II and Set III candidate-pool dimensions, Table~S7 reports all matched-rank errors, and Table~S8 gives the absolute polarizabilities and dispersion coefficients.

The complete-response reference calculations recovered the corresponding full responses within the verification tolerance for all 36 stationary molecule/method pairs.
This required spanning all $N_{\mathrm{occ}}N_{\mathrm{vir}}$ independent occupied--virtual density-matrix variations, not merely including that number of potentially dependent functions.
The density responses sampled at three nonzero frequencies completed the response basis only for \ce{H2O} in this study.
For \ce{CO} and \ce{N2}, they gave a basis of dimension 196 compared with 273 in the complete space.
For \ce{CCl4}, the corresponding dimensions were 496 and 3478.
The full \ce{CCl4}/HFsrPBE Set III basis still gave a 2.96\% directional error.
Exhausting Set III therefore did not establish complete-response convergence for the remaining molecules.

\subsection{Charge fluctuations and long-range correlation energy}
\label{subsec:correlation-results}

Table~\ref{tab:molecular-correlation-energy} presents the long-range correlation energies of \ce{H2O}, CO, and \ce{N2} obtained with ACKS2$\omega$ and the corresponding orbital-space calculations.
All comparisons in this table use the same \texttt{veryfine} molecular grid defined in HORTON.
The three energy treatments are defined in Sec.~\ref{subsec:acfdt-correlation}.
The RPAx-I calculation including the frequency-dependent exchange contribution uses Eqs.~\eqref{eq:main-longrange-exchange-hardness}--\eqref{eq:main-acfdt-condensed-energy}.
Within each response set and rank, the density and potential functions are identical for the three treatments.
Sets II and III instead compare different functions at the stated dimensions.
``Matched'' denotes equal response-basis dimensions in Sets II and III.
A full candidate pool need not span the complete response space.
Section~S9.1 and Table~S9 of the Supporting Information give all calculated energies and the changes on including the frequency-dependent exchange contribution.

\begin{table}[htbp]
\centering\small
\caption{Long-range correlation energies (millihartree) for Sets I--IV and the independent orbital reference on the HORTON \texttt{veryfine} grid, where matched Sets II/III have equal response rank and RPAx-I without the frequency-dependent exchange contribution is the static-only approximation.}
\label{tab:molecular-correlation-energy}
\begin{tabular}{llrrrr}
\toprule
\multirow{3}{*}{Molecule} & \multirow{3}{*}{Response space} & \multirow{3}{*}{$N_r/N_{\mathrm{ov}}$} & \multirow{3}{*}{dRPA-I} & \multicolumn{2}{c}{RPAx-I}\\
\cmidrule(lr){5-6}
 & & & & \makecell{Without frequency-\\dependent exchange} & \makecell{With frequency-\\dependent exchange}\\
\midrule
\multirow{7}{*}{\ce{H2O}} & Set I (original) & $74/180$ & -5.010987 & 9.819947 & -10.377910 \\
 & Set II (80\%) & $159/180$ & -5.108465 & 9.833116 & -5.173546 \\
 & Set II (full pool) & $180/180$ & -5.110122 & 9.835446 & -5.134731 \\
 & Set III (matched) & $159/180$ & -5.108936 & 9.835061 & -5.213633 \\
 & Set III (full pool) & $180/180$ & -5.110122 & 9.835446 & -5.134731 \\
 & Set IV & $180/180$ & -5.110122 & 9.835446 & -5.134731 \\
 & Orbital reference & $180/180$ & -5.110122 & 9.835446 & -5.134731 \\
\addlinespace
\multirow{8}{*}{\ce{CO}} & Set I (original) & $49/273$ & -9.538831 & 13.218626 & -27.161618 \\
 & Set II (80\%) & $169/273$ & -9.821255 & 13.627205 & -11.518145 \\
 & Set II (matched) & $196/273$ & -9.871295 & 13.734608 & -11.200189 \\
 & Set II (full pool) & $199/273$ & -9.880875 & 13.756813 & -11.174512 \\
 & Set III (matched) & $169/273$ & -9.786302 & 13.540103 & -11.585552 \\
 & Set III (full pool) & $196/273$ & -9.804222 & 13.580846 & -11.350170 \\
 & Set IV & $273/273$ & -9.880875 & 13.752828 & -10.694835 \\
 & Orbital reference & $273/273$ & -9.880875 & 13.752828 & -10.694835 \\
\addlinespace
\multirow{8}{*}{\ce{N2}} & Set I (original) & $49/273$ & -10.646546 & 12.049674 & -25.018835 \\
 & Set II (80\%) & $169/273$ & -11.114927 & 12.196438 & -13.116652 \\
 & Set II (matched) & $196/273$ & -11.182647 & 12.323583 & -12.639242 \\
 & Set II (full pool) & $199/273$ & -11.193714 & 12.347864 & -12.586654 \\
 & Set III (matched) & $169/273$ & -11.072451 & 12.142031 & -12.805415 \\
 & Set III (full pool) & $196/273$ & -11.111202 & 12.206111 & -12.596644 \\
 & Set IV & $273/273$ & -11.193706 & 12.346822 & -12.284763 \\
 & Orbital reference & $273/273$ & -11.193706 & 12.346822 & -12.284763 \\
\bottomrule
\end{tabular}
\end{table}
 
The complete ACKS2 representation reproduces the orbital-space energy within $10^{-9}$ hartree for each treatment on all three grids examined.
Table~S10 of the Supporting Information compares the Set IV ACKS2 energies with the independent orbital-space results.
On the \texttt{veryfine} grid, the orbital RPAx-I energies are $-5.134731$, $-10.694835$, and $-12.284763$ millihartree for \ce{H2O}, \ce{CO}, and \ce{N2}, respectively.
The complete-response calculation uses orbital-pair density matrices and their dual potentials in the same ACKS2 energy expression.
This agreement confirms the connection between the two representations for the selected long-range approximation.
Each reduced-set result is compared with its corresponding orbital reference.
Direct RPA and RPAx-I are different approximations to the correlation energy.

The frequency-dependent exchange contribution has a substantial effect on these energies.
At the 80\% Set II selection, including this term changes the energy by $-15.007$, $-25.145$, and $-25.313$ millihartree for \ce{H2O}, CO, and \ce{N2}, respectively, and changes its sign relative to the static-exchange-only calculation.
Positive static-only energies also occur in the complete response space.
Thus, adding more density functions does not replace the frequency-dependent exchange term.
Here, the static-only and full exchange treatments refer to alternative response approximations.
Their energy difference is not a separately additive component of the correlation energy.

\begin{figure}[htbp]
\centering
\includegraphics[width=\linewidth]{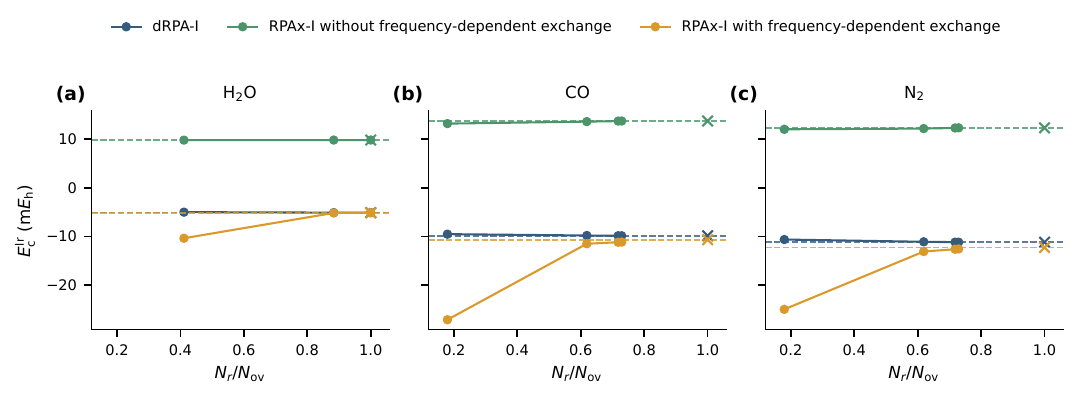}
\caption{Long-range correlation energies for \ce{H2O} (a), CO (b), and \ce{N2} (c) as the local response space is enlarged from Set I through Set II on the \texttt{veryfine} grid.
The horizontal axis gives the fraction of the complete occupied--virtual response dimension included in each response set.
Dashed lines and crosses mark the Set IV reference for each treatment.
RPAx-I without the frequency-dependent exchange contribution is the static-only approximation.}
\label{fig:molecular-energy-convergence}
\end{figure}

Set I overestimates the magnitude of the exchange-containing energy by factors of 2.021, 2.540, and 2.037 for \ce{H2O}, CO, and \ce{N2}, respectively.
Adding local radial functions markedly improves the agreement, as shown in Fig.~\ref{fig:molecular-energy-convergence}.
The original atomic multipoles therefore provide an insufficient representation for these correlation energies, even when the exchange contribution is included.
The finite-space response approximation also matters.
Inverting the represented bare independent-particle response and projecting its inverse are different operations [Eq.~(S39) in the Supporting Information].
The independent normalization check and projection comparison are given in Section~S9.2 and Table~S11 of the Supporting Information.

At the 80\% Set II selection, the absolute relative errors are 0.76\%, 7.70\%, and 6.77\% for \ce{H2O}, CO, and \ce{N2}.
The full local basis for \ce{H2O} spans all 180 occupied--virtual density-matrix variations and recovers the orbital-space result.
The full local bases for \ce{CO} and \ce{N2} have dimension 199 compared with 273 in the complete orbital space, with remaining errors of 4.49\% and 2.46\%, respectively.
Correlation energy depends on interactions between the represented density fluctuations, whereas dipole polarizability probes their response to a uniform electric field.
A response set that describes the dipole polarizability accurately may therefore require further enlargement to describe the energy.

The comparison with density-first enrichment is molecule dependent.
At the same ranks as the 80\% Set II selection, Set III gives energy errors of 1.54\%, 8.33\%, and 4.24\% for \ce{H2O}, CO, and \ce{N2}.
It improves the result for \ce{N2}, but not for \ce{H2O} or CO.
The complete Set III basis of dimension 180 for \ce{H2O} recovers the orbital-space energy.
For \ce{CO} and \ce{N2}, Set III has dimension 196 and gives errors of 6.13\% and 2.54\%, compared with 4.73\% and 2.89\% for Set II at the same rank.
Thus, the number of functions and the electronic rearrangements that they describe both affect the energy.
Neither construction is uniformly preferable at a finite rank.

On finer molecular grids, the complete ACKS2 and orbital energies remain in agreement, and successive complete-space energies change by less than $1.4\times10^{-8}$ hartree.
Rebuilding the reduced response sets on those grids produces larger energy changes, particularly for \ce{N2}.
The selected functions can change because their numerical matrix elements determine which nearly dependent density variations are kept.
The reported reduced-set energies therefore refer to the \texttt{veryfine} grid, not a grid-extrapolated limit.
Section~S9.3 and Tables~S12 and S13 of the Supporting Information give the grid comparisons and the much smaller frequency and coupling-strength quadrature errors, respectively.

\begin{figure}[htbp]
\centering
\includegraphics[width=\linewidth]{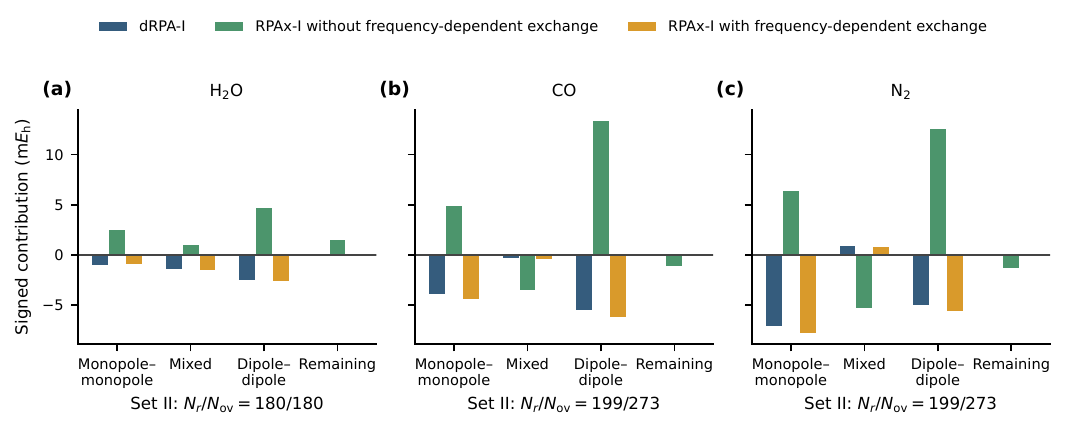}
\caption{Signed full-pool Set II energy contributions for \ce{H2O} (a), CO (b), and \ce{N2} (c) in the specified local moment representation on the \texttt{veryfine} grid.
The four groups follow Eqs.~\eqref{eq:main-correlation-multipole-parts} and~\eqref{eq:main-dipole-remainder-split} and sum to the finite-basis ACKS2 total, not the independent orbital-reference total.
Table~S14 of the Supporting Information lists the contributions and the separate finite-basis errors.}
\label{fig:molecular-energy-decomposition}
\end{figure}

Figure~\ref{fig:molecular-energy-decomposition} separates the full-pool Set II energies into monopole, mixed, dipole--dipole, and remaining contributions.
Section~S9.4 of the Supporting Information describes the moment representation and its numerical convergence.

These terms describe parts of the same coupled response, as in the distributed-polarizability analysis.\cite{Cheng2023}
The monopole group includes both net atomic-population moments and radial charge-shape moments, so it is broader than net interatomic charge flow alone.
With the frequency-dependent exchange contribution included, the monopole--monopole contributions are $-0.951737$, $-4.374551$, and $-7.789091$ millihartree for \ce{H2O}, CO, and \ce{N2}, respectively.
The dipole--dipole contributions are $-2.617883$, $-6.212972$, and $-5.630099$ millihartree.
Within this representation, the monopole contribution is smaller than the dipole contribution for \ce{H2O} and CO, but larger for \ce{N2}.
Both kinds of fluctuation contribute appreciably to the long-range correlation energy.

The remaining contributions are $-0.051147$, $-0.167330$, and $+0.003134$ millihartree for \ce{H2O}, CO, and \ce{N2}, respectively.
Their signs reflect couplings within the full response, rather than energies obtained by switching off other fluctuations.
Together with the mixed contributions, the four groups reproduce the ACKS2 totals $-5.134731$, $-11.174512$, and $-12.586654$ millihartree within $10^{-9}$ hartree.
The Set II moment representation for \ce{H2O} spans all 180 occupied--virtual density-matrix variations.
The \ce{CO} and \ce{N2} representations have dimension 199 compared with 273 in the complete orbital space.
The ACKS2-minus-orbital energy differences are negligible for \ce{H2O} and $-0.479678$ and $-0.301891$ millihartree for CO and \ce{N2}.
These differences are reported separately from the multipole contributions because the latter sum to the energy of the represented response space.

The individual contributions depend on the chosen atomic representation.\cite{Cheng2023}
The \ce{H2O} comparisons in Tables~S15 and S16 of the Supporting Information distinguish the effect of rebuilding that representation on a different grid from numerical integration with a fixed set of moment operators.
The decomposition provides a way to examine charge and local polarization contributions within a specified model.
The total energy remains the quantity compared with the orbital-space reference.
 \section{Discussion}
\label{sec:discussion}
\subsection{Electronic interactions and flexibility of the response}

The results show that molecular polarization depends both on the interactions included in the working equations and on the electronic rearrangements allowed by the response functions.
With the independent-particle auxiliary response used here, HF, B3LYP, and HFsrPBE require the additional frequency-dependent exchange contribution alongside the static exchange hardness.
Enlarging the response set improves its description of the induced density, but does not replace this contribution to the interaction.
For PBE, which contains no exact exchange, the additional exchange correction vanishes, but enlarging the response basis still improves the results.
The number of TS27 molecules meeting both 1\% criteria increases from 12 with Set I to 24 with Set II at $\zeta=0.80$ and to all 27 with the full local basis (Table~\ref{tab:full-basis-transferability}).
Thus, the benefit of additional response functions is not restricted to methods containing exact exchange.

The two enrichment approaches provide different ways to describe changes in the electron distribution.
Set II adds radial flexibility to the local perturbing potentials and then constructs the corresponding density functions.
Set III adds density responses directly, then constructs biorthogonal dual potential operators using the static independent-particle response, as specified in Eq.~\eqref{eq:main-density-first-pair}.
This construction reduces the maximum directional polarizability error for all nine local-pool targets with each exchange-containing method, although the absolute $C_6$ error improves in only five of the 27 comparisons.
This difference follows from the properties being compared.
The maximum directional error measures the largest relative deviation of the tensor, whereas $C_6$ integrates the squared isotropic polarizability over frequency.
Better agreement for one quantity need not give better agreement for the other.

Using the same enrichment rule for every molecule avoids adjusting its parameters to individual results.
The resulting density and potential functions nevertheless depend on the molecular electronic structure and AIM partition.
As in the earlier ACKS2$\omega$ study,\cite{Cheng2022} calculating these ingredients from the reference wavefunction tests the response formulation.
Transferring a fixed set of force-field parameters to new environments is a separate objective.
The present parameter construction still requires electronic-structure calculations.

\subsection{Chemical interpretation of the molecular properties}

Molecular polarization includes both charge flow between atomic regions and local polarization within them.
The distributed response describes these contributions and their coupling, while their sum gives the molecular dipole polarizability.\cite{Cheng2023}
The isotropic mean and the anisotropy emphasize different aspects of this response.
In particular, anisotropy depends on differences between tensor components and can remain sensitive when the mean polarizability is already reproduced well.

The local moment representation similarly separates the correlation energy into monopole, mixed, dipole--dipole, and remaining terms.
The non-negligible monopole and mixed terms in \ce{H2O}, CO, and \ce{N2} show why local dipoles alone do not describe all contributions in this representation.
For the original atomic population moments, the terms charge fluctuations and charge flow refer to the same monopole response.
The enriched Set II monopole group also contains radial charge-shape fluctuations and should therefore not be identified entirely with net interatomic charge flow.
The assigned contributions refer to the stated moment operators and their pairing.
Generalized Set III functions require a separately specified map before the same atomic interpretation can be applied.

\subsection{From molecular polarization to correlation energy}

The long-range correlation calculations connect ACKS2$\omega$ to an energy derived from the coupled electronic fluctuations.
At complete response rank, the ACKS2 and independent orbital-space calculations agree for the same HFsrPBE reference and dRPA-I or RPAx-I approximation.
The energy evaluation therefore maintains the connection to the orbital response while using ACKS2 response and hardness parameters.
The orbital values are references for the effect of atom condensation, not exact many-electron correlation energies.\cite{Toulouse2010,Angyan2011}

The reduced response sets show that correlation energies place different demands on the representation than dipole polarizabilities.
Set I considerably overestimates the exchange-containing energy magnitudes.
Both enrichment strategies improve their description, but the local and density-first sets need not give the same ordering of errors at equal rank.
A complete set recovers the reference, whereas exhausting a chosen family of candidate functions does not necessarily complete the response space.

The main energy comparisons use identical molecular grids and electronic references for the condensed and orbital calculations.
Separate finer-grid calculations examine how the numerical construction of the response functions changes the results.
The complete-space energies are stable over the tested grids, while some reduced sets change appreciably when rebuilt.
Thus the finite-set comparisons describe the stated construction on the stated grid.
The integration and response-set effects are reported separately in Tables~S12--S13 and S15--S16 of the Supporting Information.
This distinction is also relevant to individual energy contributions.
Holding a chemical representation fixed and rebuilding it on another grid are different comparisons.

In extended systems, the gap, dimensionality, and separation affect the range of charge fluctuations and the asymptotic dispersion interaction.\cite{Dobson2006,Misquitta2010,Misquitta2014}
The molecular results do not determine these extended-system asymptotics.
Testing them requires calculations beyond the molecular set studied here.
 \section{Conclusions}
\label{sec:summary}
We have applied the coupled density--potential formulation of ACKS2$\omega$ to molecular references containing exact exchange, including HF, B3LYP, and HFsrPBE.
With the independent-particle auxiliary response used here, exchange contributes both a static hardness and an additional frequency-dependent exchange contribution.
The latter vanishes in the static limit but substantially affects frequency-dependent polarizabilities, dispersion coefficients, and long-range correlation energies.

The response-set comparisons distinguish the treatment of exchange from the flexibility of the induced density.
Local radial enrichment improves the TS27 response, while density-first enrichment reduces the maximum directional error for the nine local-pool targets with all three exchange-containing methods.
The corresponding $C_6$ errors improve in only five of the 27 matched-rank comparisons.
Thus the two constructions are complementary, and neither provides a uniformly more accurate finite response set.
Complete-response calculations recover the corresponding orbital-space results.

For \ce{H2O}, CO, and \ce{N2}, long-range correlation energies were calculated from ACKS2 response and hardness parameters and compared with independent orbital-space ACFDT integrals.
At complete response rank, the two approaches agree for the same HFsrPBE reference and long-range dRPA-I or RPAx-I approximation.
Reduced sets still show representation errors and a dependence on the molecular grid used to construct them, as quantified separately from frequency and coupling-strength integration errors.

The monopole, mixed, dipole--dipole, and remaining contributions reconstruct the finite-set ACKS2 energy in the chosen atomic moment representation.
For the original atomic population moments, the monopole terms describe charge fluctuations, equivalently charge flow.
In the enriched Set II representation, the monopole group additionally contains radial charge-shape fluctuations.
The resulting monopole and mixed contributions are appreciable alongside local dipolar polarization.
ACKS2$\omega$ can therefore evaluate both the long-range correlation energy and its atom-condensed contributions from the same response and hardness parameters.
 
\section*{Data and Software Availability}
The supporting data and code are available from the corresponding author on reasonable request.

\section*{Supporting Information}
Supplementary details of the ACKS2 response bases, exchange contribution, and correlation-energy expressions, electronic-reference preparation and numerical protocols, local response-set selection, per-case TS27 results and numerical tests, density-first response comparisons, and molecular correlation energies and multipole contributions (PDF).

\section*{Competing Financial Interests}
The author declares no competing financial interest.

\section*{Acknowledgments}
This work was supported by the National Natural Science Foundation of China (T2600851).

\bibliography{article}

\clearpage
\begingroup
\nolinenumbers
\begin{center}
\textbf{For Table of Contents Only}
\par\medskip
\includegraphics[width=3.25in,height=1.75in,keepaspectratio]{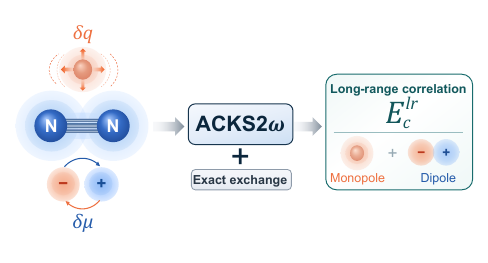}
\end{center}
\endgroup
\end{document}